\documentclass[twocolumn]{aastex631}

\newcommand{\feii}{[Fe \textsc{ii}]}
\newcommand{\pab}{Pa$\beta$}

\newcommand{\ha}{H$\alpha$}
\newcommand{\hb}{H$\beta$}

\newcommand{\oiii}{[O \textsc{iii}]}
\newcommand{\oii}{[O \textsc{ii}]}

\newcommand{\nii}{[N \textsc{ii}]}
\newcommand{\neiii}{[Ne \textsc{iii}]}
\newcommand{\sii}{[S \textsc{ii}]}
\newcommand{\hii}{H \textsc{ii}}

\usepackage{tabularx}
\usepackage{amsmath}
\usepackage{amstext}
\usepackage{apjfonts}
\usepackage{graphicx}
\usepackage{tablefootnote}
\usepackage{amssymb}
\usepackage{latexsym}
\usepackage{longtable}

\accepted{June 9, 2025}

\submitjournal{The Astrophysical Journal}

\shorttitle{Shocks Trace Feedback in Starburst LIRG VV 114}
\shortauthors{Kader et al.}
\graphicspath{{./}{figures/}}
\begin{document}

\title{Shockingly Effective: Cluster Winds as Engines of Feedback in Starburst Galaxy VV 114}

\author[0000-0002-6650-3757]{Justin A. Kader}
\affiliation{Department of Physics and Astronomy, University of California, Irvine, 4129 Frederick Reines Hall, Irvine, CA 92697, USA}

\author[0000-0002-1912-0024]{Vivian U}
\affiliation{IPAC, California Institute of Technology, 1200 E. California Blvd., Pasadena, CA 91125, USA}
\affiliation{Department of Physics and Astronomy, University of California, Irvine, 4129 Frederick Reines Hall, Irvine, CA 92697, USA}

\author[0000-0002-5807-5078]{Jeffrey A. Rich}
\affiliation{The Observatories of the Carnegie Institution for Science, 813 Santa Barbara Street, Pasadena, CA 91101, USA}

\author[0000-0002-6570-9446]{Marina Bianchin}
\affiliation{Department of Physics and Astronomy, University of California, Irvine, 4129 Frederick Reines Hall, Irvine, CA 92697, USA}

\author[0000-0002-1000-6081]{Sean T. Linden}
\affiliation{Department of Astronomy, University of Arizona, 1114 E Lowell Road, Tucson, AZ 85719, USA}

\author[0000-0001-7421-2944]{Anne M. Medling}
\affiliation{Department of Physics \& Astronomy and Ritter Astrophysical Research Center, University of Toledo, Toledo, OH 43606, USA}

\author[0000-0003-0699-6083]{Tanio Diaz-Santos}
\affiliation{Institute of Astrophysics, Foundation for Research and Technology-Hellas (FORTH), Heraklion, 70013, Greece}
\affiliation{School of Sciences, European University Cyprus, Diogenes street, Engomi, 1516 Nicosia, Cyprus}

\author[0000-0003-3474-1125]{George C. Privon}
\affiliation{National Radio Astronomy Observatory, 520 Edgemont Road, Charlottesville, VA 22903, USA}
\affiliation{Department of Astronomy, University of Virginia, 530 McCormick Road, Charlottesville, VA 22904, USA}
\affiliation{Department of Astronomy, University of Florida, P.O. Box 112055, Gainesville, FL 32611, USA}

\author[0000-0003-2064-4105]
{Rosalie McGurk}
\affiliation{W. M. Keck Observatory, 65-1120 Mamalahoa Hwy. Kamuela, HI 96743, USA}

\author[0000-0003-3498-2973]{Lee Armus}
\affiliation{IPAC, California Institute of Technology, 1200 E. California Blvd., Pasadena, CA 91125, USA}

\author[0000-0003-0057-8892]{Loreto Barcos-M\~{u}noz}
\affiliation{National Radio Astronomy Observatory, 520 Edgemont Road, Charlottesville, VA 22903, USA}
\affiliation{Department of Astronomy, University of Virginia, 530 McCormick Road, Charlottesville, VA 22904, USA}

\author[0000-0003-4693-6157]{Gabriela Canalizo}
\affiliation{Department of Physics and Astronomy, University of California, Riverside, 900 University Avenue, Riverside, CA 92521, USA}

\author[0000-0002-2688-1956]{Vassilis Charmandaris}
\affiliation{Institute of Astrophysics, Foundation for Research and Technology-Hellas (FORTH), Heraklion, 70013, Greece}
\affiliation{School of Sciences, European University Cyprus, Diogenes street, Engomi, 1516 Nicosia, Cyprus}
\affiliation{Department of Physics, University of Crete, Heraklion, 71003, Greece}

\author[0000-0003-2638-1334]{Aaron S. Evans}
\affiliation{Department of Astronomy, University of Virginia, 530 McCormick Road, Charlottesville, VA 22904, USA}
\affiliation{National Radio Astronomy Observatory, 520 Edgemont Road, Charlottesville, VA 22903, USA}

\author[0000-0002-1158-6372]{Tianmu Gao}
\affiliation{Research School of Astronomy and Astrophysics, Australian National University, Weston Creek, ACT 2611, Australia}
\affiliation{ARC Centre of Excellence for All Sky Astrophysics in 3 Dimensions (ASTRO 3D), Canberra, ACT 2611, Australia}

\author[0000-0001-6028-8059]{Justin Howell}
\affiliation{IPAC, California Institute of Technology, 1200 E. California Blvd., Pasadena, CA 91125, USA}

\author[0000-0003-4268-0393]{Hanae Inami}
\affiliation{Hiroshima Astrophysical Science Center, Hiroshima University, 1-3-1 Kagamiyama, Higashi-Hiroshima, Hiroshima 739–8526, Japan}

\author[0000-0001-8490-6632]{Thomas Lai}
\affiliation{IPAC, California Institute of Technology, 1200 E. California Blvd., Pasadena, CA 91125, USA}

\author[0000-0003-3917-6460]{Kirsten L. Larson}
\affiliation{AURA for the European Space Agency (ESA), Space Telescope Science Institute, 3700 San Martin Drive, Baltimore, MD 21218, USA}

\author[0000-0001-6919-1237]{Matthew A. Malkan}
\affiliation{Department of Physics \& Astronomy, University of California, Los Angeles, Los Angeles, CA 90095-1547, USA}

\author[0000-0003-4286-4475]{Mar\'ia S\'anchez-Garc\'ia}
\affil{Institute of Astrophysics, Foundation for Research and
Technology-Hellas (FORTH), Heraklion, 70013, Greece}

\author{Christopher D. Martin}
\affiliation{Cahill Center for Astronomy and Astrophysics, California Institute of Technology, 1200 E. California Blvd. MC 249-17, Pasadena, CA 91125, USA}

\author{Mateusz Matuszewski}
\affiliation{Cahill Center for Astronomy and Astrophysics, California Institute of Technology, 1200 E. California Blvd. MC 249-17, Pasadena, CA 91125, USA}

\author[0000-0003-0682-5436]{Claire E. Max}
\affiliation{Department of Astronomy \& Astrophysics, University of California, Santa Cruz, CA 95064, USA}

\author[0000-0002-8204-8619]{Joseph M. Mazzarella}
\affiliation{IPAC, California Institute of Technology, 1200 E. California Blvd., Pasadena, CA 91125, USA}

\author[0000-0002-0466-1119]{James D. Neill}
\affiliation{Cahill Center for Astronomy and Astrophysics, California Institute of Technology, 1200 E. California Blvd. MC 249-17, Pasadena, CA 91125, USA}

\author[0000-0001-5847-7934]{Nikolaus Z. Prusinski}
\affiliation{Cahill Center for Astronomy and Astrophysics, California Institute of Technology, 1200 E. California Blvd. MC 249-17, Pasadena, CA 91125, USA}

\author[0000-0002-0164-8795]{Raymond Remigio}
\affiliation{Department of Physics and Astronomy, University of California, Irvine, 4129 Frederick Reines Hall, Irvine, CA 92697, USA}

\author[0000-0002-1233-9998]{David Sanders}
\affiliation{Institute for Astronomy, University of Hawaii, 2680 Woodlawn Drive, Honolulu, HI 96822, USA}

\author[0000-0002-3139-3041]{Yiqing Song}
\affiliation{European Southern Observatory, Alonso de Córdova, 3107, Vitacura, Santiago, 763-0355, Chile}
\affiliation{Joint ALMA Observatory, Alonso de Córdova, 3107, Vitacura, Santiago, 763-0355, Chile}

\author[0000-0002-2596-8531]{Sabrina Stierwalt}
\affiliation{Occidental College, Physics Department, 1600 Campus Road, Los Angeles, CA 90042, USA}

\author[0000-0001-7291-0087]{Jason Surace}
\affiliation{IPAC, California Institute of Technology, 1200 E. California Blvd., Pasadena, CA 91125, USA}



\begin{abstract}

We present high-resolution Keck Cosmic Web Imager (KCWI) and MUSE IFU spectroscopy of VV 114, a local infrared-luminous merger undergoing a vigorous starburst and showing evidence of galactic-scale feedback. The high-resolution data allow for spectral deblending of the optical emission lines and reveal a broad emission line component ($\sigma_{\rm{broad}} \sim$~100--300 km s$^{-1}$) with line ratios and kinematics consistent with a mixture of ionization by stars and radiative shocks. The shock fraction (percent ionization due to shocks) in the high velocity gas is anticorrelated with projected surface number density of resolved star clusters, and we find radial density profiles around clusters are well fit by models of adiabatically expanding cluster winds driven by massive stellar winds and supernovae (SNe). The total kinetic power estimated from the cluster wind models matches the wind+SNe mechanical energy deposition rate estimated from the soft band X-ray luminosity, indicating that at least 70\% of the shock luminosity in the galaxy is driven by the star clusters. \emph{Hubble Space Telescope} narrow band near-infrared imaging reveals embedded shocks in the dust-buried infrared nucleus of VV 114E. Most of the shocked gas is blueshifted with respect to the quiescent medium, and there is a close spatial correspondence between the shock map and the \emph{Chandra} soft band X-ray image, implying the presence of a galactic superwind. The energy budget of the superwind is in close agreement with the total kinetic power of the cluster winds, confirming the superwind is driven by the starburst.

\end{abstract}

\keywords{VV~114, Interstellar shocks, Luminous Infrared Galaxies, Galactic feedback}


\section{Introduction} \label{sec:intro}

Galaxy mergers serve as extreme laboratories for studying feedback—the process by which energy and momentum input from star formation and active galactic nucleus (AGN) activity regulate the growth of galactic stellar mass and supermassive holes. While cosmological simulations routinely invoke feedback to explain galaxy properties both on the individual and population levels, its implementation remains highly idealized, particularly on sub-resolution scales where stellar winds and supernovae interact with the multiphase ISM \citep{schaye15, naabandostriker17, vogelsberger20}. These theoretical limitations create a pressing need for empirical constraints, yet observational studies face their own challenges. Although galactic outflows are commonly detected in starbursts and mergers, their driving mechanisms—and ultimate impact on star formation—remain poorly understood. Radiative shocks, for instance, may either suppress star formation by disrupting molecular clouds \citep{hopkins13} or enhance it through gas compression and inflows \citep{elmegreen78, kinoshita21, rich11}. Such ambiguities persist because most observations lack the resolution to connect galaxy-scale outflows to their originating sources (e.g., individual star clusters) or trace their local interaction with the ISM.

This gap in understanding stems largely from the inherently multiscale nature of feedback, which operates across spatial scales from star-forming regions (tens of pc) to galaxy-wide outflows (several to tens of kpc). Current facilities cannot simultaneously resolve these scales in distant galaxies, making local luminous and ultra-luminous infrared galaxies (LIRGs/ULIRGs; $L_{\rm{IR}} > 10^{11}~L_{\odot}$ \citep{Soifer87}) indispensable laboratories. These extreme starbursts host powerful galactic winds and pervasive shocks \citep{veilleux05, rich11, westmoquette12, rich15}, while their proximity enables observations at the critical scales needed to dissect feedback mechanisms. Studying how energy propagates from individual sources through the ISM in these systems provides fundamental insights into how energetic processes govern galaxy evolution.

The transformative impact of AGN-powered outflows and galactic superwinds on their host galaxies is evident in their localized effects on the ISM, with shocks being one of the most easily observed signposts of feedback. Shocks are present whenever matter propagates through a medium at a rate faster than the local sound speed, and often accompany galactic outflows, jet-interstellar medium (ISM) interactions, supernovae (SNe), and large-scale tidally-induced gas motions in mergers (e.g., \citealt{saito17}.) While shocks can suppress star formation by disrupting giant molecular clouds (GMCs) over short timescales ($\lesssim$1 Myr; \citealt{hopkins13}), they may also work to promote star formation over longer timescales by radiatively dissipating the kinetic energy of ambient gas, causing it to fall into the potential well of the galaxy (e.g., \citealt{rich11}), or on local scales via gas compression (e.g., \citealt{shinichi21}, \citealt{humire25}.) (U)LIRGs are ideal testbeds for estimating the energetics and timescales involved in AGN- and starburst-powered feedback through shocks which propagate through the gas-rich ISM. In this paper, high-resolution, wide areal coverage optical spectroscopic and near-IR photometric observations of the shocked ISM are used to constrain the energetics, timescales, and feedback effect of a galactic-scale starburst.

The LIRG ($L_{\rm{IR}}$[8 -- 1000 $\mu$m] = 4.5$\times$10$^{11} L_{\odot}$) VV 114 (= IRAS F01053-1764 = IC 1623 = Arp 236) is a nearby ($D_{\rm{L}}$ = 84 Mpc, 1$\arcsec$ = 410 pc) mid-stage merger in the Great Observatories All-sky LIRG Survey (GOALS; \citealt{Armus09}), consisting of two progenitor galaxies with a projected nuclear separation of approximately 8 kpc \citep{Evans22}. The merger is gas-rich, with a total molecular gas mass of $M_{\rm{H_2}}$ = 5.1$\times$10$^{10}$ $M_{\odot}$ \citep{Yun94} and total neutral gas mass of $M_{\rm{H I}}$ = 7.2$\times$10$^9$ $M_{\odot}$ \citep{vanDriel01}, and with massive stellar ($M_{*}$ = 1.62$\times$10$^{11}$ $M_{\odot}$; \citealt{Howell10}) and dust ($M_{\rm{dust}}$ = 1.2$\times$10$^8$ $M_{\odot}$; \citealt{Frayer99}) components. The eastern component (VV 114E) is heavily obscured by dust, rendering it invisible at UV wavelengths, but beyond $\sim$1 $\mu$m, VV 114E becomes the dominant luminosity source of the merger \citep{Knop94,Doyon95,Scoville00,Soifer01,LeFloch02,charmandaris04,Evans22}. The total IR-to-UV flux ratio of VV 114 is 11.2 \citep{Howell10}, indicating that most of the energy sources in the system are buried behind dust in the eastern component. Mostly due to the long wavelength emission from the eastern component, VV 114 is one of the brightest objects in the IRAS Bright Galaxy Sample \citep{Soifer87}. Keck MERLIN observations at 12.5 $\mu$m resolve the nucleus of VV 114E into two bright cores, one to the NE and the other to the SW, separated by $\sim$1.$\arcsec$6 (= 630 pc; \citep{Soifer01,Evans22}, possibly indicating that VV 114E is itself a merger remnant \citep{rich11}. James Webb Space Telescope (\emph{JWST}) imaging reveals that the NE and SW cores have mid-IR colors consistent with a dust-buried starburst and a heavily obscured AGN, respectively \citep{Evans22}. In line with this picture, \emph{JWST} near- and mid-IR spectroscopy shows properties ro-vibrational emission signatures of molecular gas toward the AGN are inconsistent with those expected in starbursts \citep{buiten24}. \emph{JWST} mid-IR spectroscopy also reveals deep absorption due to aliphatic hydrocarbons and amorphous silicates in the NE core and low polycyclic aromatic hydrocarbon (PAH) equivalent widths in the SW core, consistent with a heavily obscured starburst and AGN, respectively \citep{Rich23}. The nucleus of VV 114E also shows elevated \feii/Pf$\alpha$ line intensity ratios, broadening and multiple kinematic components in the mid-IR H$_2$ and forbidden lines, suggestive of shocks and outflows associated with and extending by $\sim$1 kpc beyond the bright IR cores \citep{Rich23}.

The western component (VV\,114W) is bright at UV and optical wavelengths \citep{Knop94,Goldader02}, hosting numerous UV-bright massive star clusters \citep{linden21}. Although both VV 114E and VV 114W feature hard-band (3-8 keV) X-ray emission (brighter and more compact in VV 114E), almost all of the soft-band (0.3-2.0 keV) X-ray emission from the system is detected towardsr and around VV\,114W \citep{Grimes06}. \emph{HST} STIS imaging shows that VV\,114W has a large far-UV luminosity ($L_{\rm{FUV}} = 2.2\times 10^{10} L_{\odot}$), comparable to high redshift Lyman break galaxies \citep{Heckman05}. \emph{FUSE} far-UV spectroscopy reveals strong, broad interstellar absorption lines with a pronounced blueshifted component toward and around VV\,114W, while WiFeS optical integral field unit (IFU) spectroscopic observations of VV 114 show elevated diagnostic line ratios across the merger \citep{Grimes06, rich11}. These observations imply strong feedback activity in the form of outflows and shocks in VV 114.

\begin{figure}[h!]
    \centering
    \includegraphics[width=0.47\textwidth]{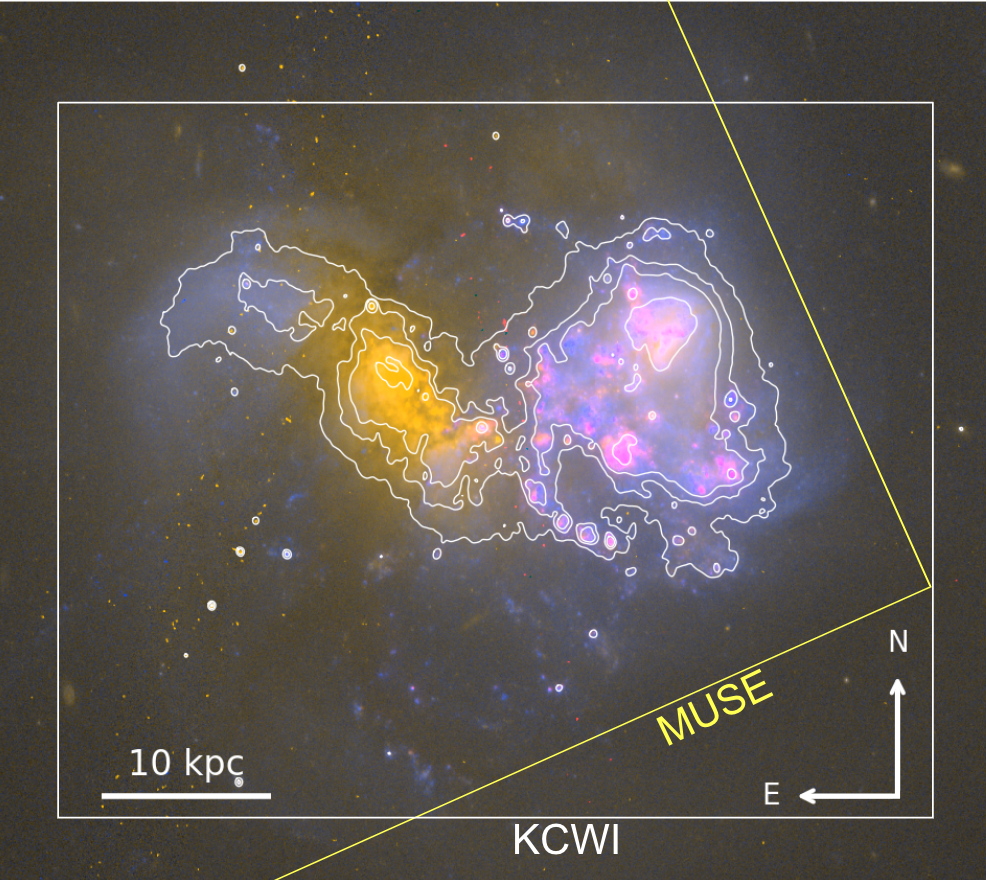}
    \caption{\emph{HST} ACS/WFC three-channel optical image of VV 114. Red is the F660N image isolating H$\alpha$ emission, green is the F814W ($\sim$I band) image and blue is the F435W ($\sim$B band) image. The tilted yellow and upright white boxes are the MUSE and KCWI fields of view, respectively. Isointensity levels of the F814W image are shown as white contours, which will also appear in all maps presented hereafter. VV\,114W appears blue with pink \hii~regions, VV 114E appears orange due to its large visual extinction. 1$\arcsec$ = 400pc, North is up and East is to the left. There is significant overlap between the IFU datasets, which covers the full merger system. An annotated version of this map, with the various regions of interest marked, is shown in Figure \ref{Figure_Atlas} in Appendix \ref{appendixA}.}
    \label{Figure_RGB}
\end{figure}

This paper is organized as follows. In Section \ref{sec:data_analysis}, we discuss our observations and data analysis techniques. In Section \ref{sec:Results}, we present the results of the  optical emission line spectroscopic analysis, and provide a theoretical shock mixing sequence framework to interpret the distribution of diagnostic emission line ratios. Section \ref{sec:Discussion} includes physical interpretations of the narrow and broad emission line components based on their respective velocity dispersions and diagnostic line ratio distributions. The optical spectroscopic and near-IR photometric results are discussed in the context of feedback in VV 114 on a galactic scale and within the dust-obscured nucleus of VV 114E. Evidence for a galactic superwind and its properties are also discussed in Section \ref{sec:Discussion}. A summary of the work and the main conclusions are enumerated in Section \ref{sec:Conclusions}. Throughout this work, we assume a standard cosmology of $\Omega_m$ = 0.286, $\Omega_{\Lambda}$ = 0.714, and $H_0$ = 69.3 km s$^{-1}$ Mpc$^{-1}$ based on the 13-year Wilkinson Microwave Anisotropy Probe Cosmology \citep{hinshaw13}.

\section{Data and Analysis} 
\label{sec:data_analysis}

\subsection{KCWI Optical IFU Data}
\label{subsec:data_kcwi}
Observations were conducted with the Keck Cosmic Web Imager (KCWI, \citealt{morrisey18}), a wide-field, seeing-limited integral field spectrograph optimized for high throughput in the 3500--5600~\AA~bandpass on the Keck II telescope. The KCWI observations presented here were obtained on UT 2021 Sep 13 under clear sky conditions, with subarcsecond seeing $\sim$0.\arcsec63. We employed a 9-point dither pattern, arranged in a 3 $\times$ 3 rectangular grid centered on VV~114, for contiguous coverage across a 35$\arcsec$ $\times$ 50$\arcsec$ area, plus two dedicated sky exposures. All exposures were 300 s long. Most of the central regions were sampled at least twice due to overlap in the dither pattern, resulting in effective integrations of greater than 300 s there.

The observations were made with the instrument in the BM-L configuration (blue medium resolution grating with the large slicer), for a central wavelength of 4100~\AA ~and a dispersion of 0.24~\AA ~pixel$^{-1}$. This setup results in a resolution of R$\sim$2000 ($\Delta v_0$=64 km s$^{-1}$) across the spectral range 3650--4550~\AA, covering the [O II] $\lambda\lambda$~3726, 3729, [Ne III] $\lambda$~3869, [Ne III] $\lambda$~3967, H$\delta~\lambda$~4102, and H$\gamma~\lambda$~4340~\AA ~emission lines. The setup gives a field of view of 33$\arcsec$ $\times$ 20.$\arcsec$4, with slit width-limited angular resolution (0.$\arcsec$29 sampling rate along the 1.$\arcsec$4-wide slit.) The detector was binned by a factor of 2 $\times$ 2. Bias frames, dark frames, thorium-argon calibration lamp spectra, as well as dome flats and twilight flats were collected during the observing night.

Basic reduction and wavelength solution for the spectroscopic data were performed using the official KCWI data reduction pipeline \textsc{KCWI\_DRP v1.0}\footnote{https://kcwi-drp.readthedocs.io/en/latest/}. The reduction pipeline also performs flux calibration of the science frames using observations of standard stars, and applies sky subtraction using the dedicated sky pointings to account for contamination by sky emission lines and continuum.  The instrumental contribution to the emission line widths was measured by fitting Gaussians to several emission lines with high signal-to-noise ratio (SNR) in the thorium-argon lamp exposures. The instrumental width (FWHM) for the emission lines was found to be 2~\AA.

The individual KCWI spectral cubes were cropped and co-added using the \textsc{CWITools}\footnote{https://cwitools.readthedocs.io/en/v0.8/} package \citep{osullivan20}, written in Python. The spectral axis of the KCWI cubes was cropped to remove coverage outside the 3650--4550~\AA~range, and extra padding in the x, y spatial axes, which is introduced in the official DRP. \textsc{CWITools} co-adds the individual cubes by placing all spaxels onto a common co-add grid. The minimum spatial sampling of the input is taken as the uniform spatial sampling of the output co-added cube, in this case resulting in square 0.$\arcsec$29 (= 120 pc) spaxels. In areas covered by only one pointing, pairs of spaxels in the horizontal (RA) direction contain identical 1-D spectra.

\subsection{MUSE Optical IFU Data}
\label{sec:data_muse}

The publicly-available MUSE data were collected from the ESO Science Archive Facility (Program Id 097.B-0427; PI: G. Privon). The reduced MUSE spectral cube has a spectral resolution of R$\sim$3000 ($\Delta v_0$=43 km s$^{-1}$) across the spectral range 4750--9352~\AA, covering the H$\beta$ $\lambda$~4861, [O III]~$\lambda$~4959, [O III]~$\lambda$~5007, [N II]~$\lambda$~6548, H$\alpha$~$\lambda$~6563, [N II]~$\lambda$~6583, and [S II]~$\lambda\lambda$~6716, 6731~\AA~emission lines. The square field of view is 1.$\arcmin$88 per side with a spatial sampling rate of 0.$\arcsec$2, and overlaps with most of the area covered by KCWI. The MUSE cube was cropped spatially to match the on-sky coverage of the KCWI co-added cube, and rebinned to give identically-sized spaxels.

\subsection{\emph{HST}/WFC3 Near-Infrared Imaging Data}
\label{sec:data_HST}

Near-IR images of VV 114 were collected with the \emph{Hubble Space Telescope} (\emph{HST}) Wide Field Camera 3 (WFC3) as part of a larger program (Program ID 17285; PI: V. U) to identify and characterize shocked atomic gas in (U)LIRGs using near-IR emission line intensity ratios. Images of VV 114 in the F110W broad-band filter and the F128N narrow-band filter were collected on 7 June, 2023 during a single orbit. The archival \emph{HST}/WFC3 F160W and F130N images of VV 114, taken on 23 July, 2019 (Program ID 15649; PI: R. Chandar) were obtained from the Hubble Legacy Archive.

The two emission lines targeted in the program were the \feii~$\lambda$1.26$\mu$m fine structure line of Fe$^+$ and the \pab~$\lambda$1.28$\mu$m hydrogen recombination line, since the \feii/\pab~emission line ratio is an established diagnostic of shocks in the ISM \citep{Larkin88,Riffel13,Riffel21}. \feii~is a particularly powerful shock diagnostic since in the normal ISM, iron is heavily depleted onto dust grains, whereas the passage of shocks breaks up grains, liberating iron back into the gas phase to be photoionized. To estimate the line emission from the two narrow-band images it was necessary to remove the continuum contribution which was measured from the two broad-band images. The first step in the image analysis was to register all four images to a common pixel coordinate space so that the same pixel position on two images corresponds to the same position on the sky. Image registration was done using the python package {\textsc{ASTROALIGN}}\footnote{https://astroalign.quatrope.org/en/latest/}, which identifies common asterisms in multiple images and estimates the affine transformation between them. After image registration, a 3$\sigma$ cut to the \emph{HST} counts images was applied (noise measured in several regions of the images far from the main galaxy.) Images were then converted to flux density units. Following \cite{larson20}, the continuum in each pixel of the narrow-band images is estimated by linear interpolation using the flux in the neighboring blue (F110W) and red (F160W) images, and then subtracted to yield a line flux image.

The reduction steps above resulted in line images of the \feii~and \pab~transitions, i.e., the continuum-subtracted F128N and F130N images, covering a wide field (2.$\arcmin$3 $\times$ 2.$\arcmin$1) that includes the entire galaxy with an on-sky angular sampling rate of 0.$\arcsec$126 in both detector directions. The \feii/\pab~image is the result of taking the simple ratio of the two line images and masking pixels with negative flux (the result of continuum subtraction using noisy pixels in the broad-band images.) As a final step, the \feii/\pab~image was registered to the KCWI cube so the line ratio could be compared with the optical IFU data on a pixel-by-pixel basis.

\subsection{Resolved Star Cluster Catalog}
\label{sec:data_ancillary}

Some of the results presented here are based on the resolved super star cluster (SSC) catalog from \cite{linden21}, which was constructed using resolved source photometry from \emph{HST} WFC3/UVIS F336W and ACS/WFC F814W and F435W broad-band optical imaging, as detailed in Evans et al. (in preparation). Sources identified with Source Extractor \citep{bertin96} across all three bands were considered detections. Of the 453 detected sources, clusters were defined as those with a FWHM $<$ 3 pixels, corresponding to an average size of $\sim$22 pc. The final catalog includes 180 confirmed clusters across the VV 114 merger.

\subsection{Optical Emission line Fitting and Decomposition}
\label{section:spectraldecomp}

The KCWI+MUSE observations resulted in 21,296 independent optical spectra spanning the full extent of VV~114. Gaussian functions were fit to the bright emission lines using the python spectral analysis tool \textsc{BADASS}\footnote{https://github.com/remingtonsexton/BADASS3} \citep{sexton21}. The input spectra are first corrected for cosmological redshift and the contribution from instrumental broadening, before being modeled by several physical components. The continuum and absorption features were modeled as a stellar line-of-sight velocity distribution (LOSVD), using Penalized Pixel-Fitting (\textsc{pPXF}\footnote{https://pypi.org/project/ppxf/}, \citealt{cappellari17}) which employed the simple stellar population template from the Indo-U.S. Library of Coud\'{e} Feed Stellar Spectra \citep{valdes04}, chosen to match the spectral resolution of the IFU observations.
Emission lines were fit with Gaussian functions. Further decomposition of the spectra to include Fe II emission features and an AGN power-law continuum did not improve the quality of the fit for the majority of the spaxels, so these components were not included.

The stellar continuum and emission lines were fit simultaneously in each of the individual spectra using an iterative maximum likelihood approach. The parameters of the best-fitting components and their uncertainties, e.g., stellar velocities, emission line fluxes, velocities, and dispersions are returned along with the $\chi^2$ goodness-of-fit statistic, as well as the total model spectrum. We computed the SNR for each emission line by taking the ratio of the peak flux density to the rms standard deviation in the flux density of the redward neighboring continuum.

Visual inspection reveals widespread complex emission line profiles. The non-Gaussian emission line profiles tend to be asymmetric and skewed towards the blue, however, lines with redward skew, or double-peaked profiles were also observed. Each emission line profile was fit with a single Gaussian component model and then a double Gaussian component model. The Bayesian Information Criterion (BIC) was used to evaluate whether a double component Gaussian model representation of the emission lines could be justified. The BIC is similar to the $\chi^2$ goodness-of-fit statistic, but includes a term that penalizes over-fitting:

\begin{equation}
    \textrm{BIC} = k{\rm{ln}}(n) - 2{\rm{ln}}(\hat{L}),
\end{equation}

\noindent where $\hat{L}$ is the likelihood function of the model, $k$ is the number of parameters for the model (2 for a single Gaussian and 4 for double Gaussians), and $n$ is the number of spectral points in the data set by the spectral sampling rate of the instrumental configuration. The significance of the double-component model was assessed by comparing the BIC values of the models:

\begin{equation}
    \delta_{\textrm{BIC}} = \textrm{BIC}_{\textrm{Single Gaussian}} - \textrm{BIC}_{\textrm{Double Gaussians}}.
\end{equation}

\begin{figure*}[ht!]
    \centering
    \includegraphics[width=\textwidth]{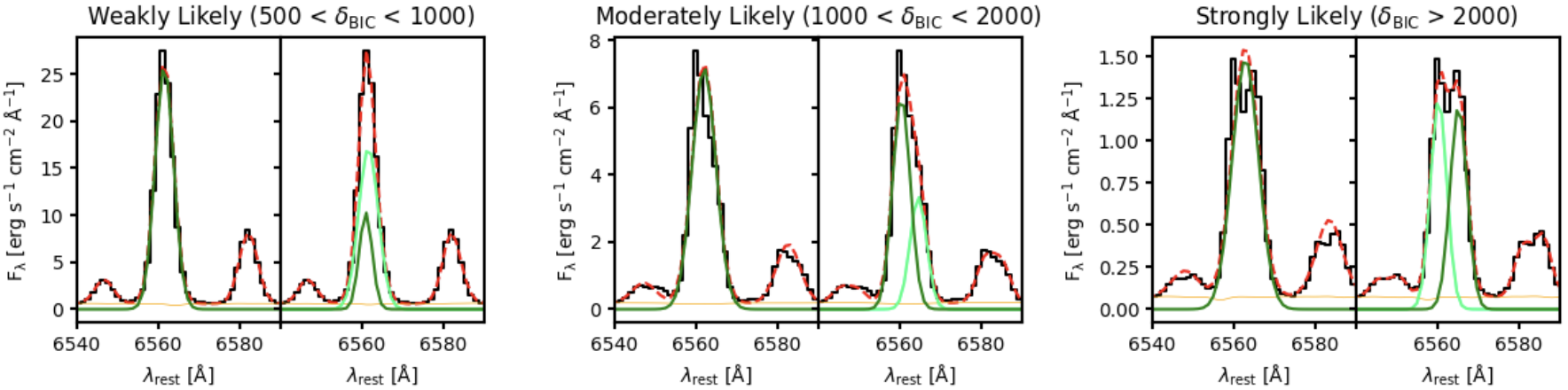}
    \caption{Single- and double-component fits to the H$\alpha$ emission line. Shown in the three panels are examples of where the statistical model comparison metric ($\delta_{\textrm{BIC}}$, see Section \ref{section:spectraldecomp}) indicates a weak (left), moderate (middle), and strong (right) likelihood that a double-component fit is the best description of the data. In the present work, emission line profiles with $\delta_{\textrm{BIC}}<1000$ are classified as single-component lines.}
    \label{Figure_LineFits}
\end{figure*}

\noindent Conventionally, a threshold of $\delta_{\textrm{BIC}}>10$ is used to rule out the hypothesis that the simpler model is the best description of the data \citep{kass95,swinbank19,avery21}. In a manner similar to \cite{rehichardtchu22}, who opted for a much larger threshold value of $\delta_{\textrm{BIC}}>500$ for spectrally resolving outflows in a star-bursting disk galaxy, for the data presented here, a threshold of $\delta_{\textrm{BIC}}>1000$ was most suitable for classifying \emph{bona fide} double-component emission lines. Spaxels were further divided into those with a weak ($500<\delta_{\textrm{BIC}}\leq~1000$), moderate ($1000<\delta_{\textrm{BIC}}\leq~2000$), and strong ($\delta_{\textrm{BIC}}>2000$) likelihood that a single-component model is inadequate to describe the emission line profiles, examples of which are shown in Figure \ref{Figure_LineFits}. In the present work, only emission lines with signal-to-noise SNR$_{\rm{line}} \geq 10$ were considered.

\section{Results}
\label{sec:Results}
This Section provides an overview of the results drawn from the analysis of the optical spectroscopic datasets. First, a statistical description of the optical emission line decomposition and fitting to the IFU data (Section \ref{sec:R1_decomposition}) is provided, followed by detailed descriptions of the individual emission line component velocity dispersion maps and excitation diagnostic diagrams (Sections \ref{sec:R2_Kinematics}--\ref{sec:R3_OpticalBPT}). Finally, in Section \ref{sec:R4_MixingSequence}, a theoretical photoionization-shock ionization mixing sequence framework is developed.

\subsection{Emission Line Decomposition Statistics}
\label{sec:R1_decomposition}

\begin{figure}[h!]
    \centering
    \includegraphics[width=0.45\textwidth]{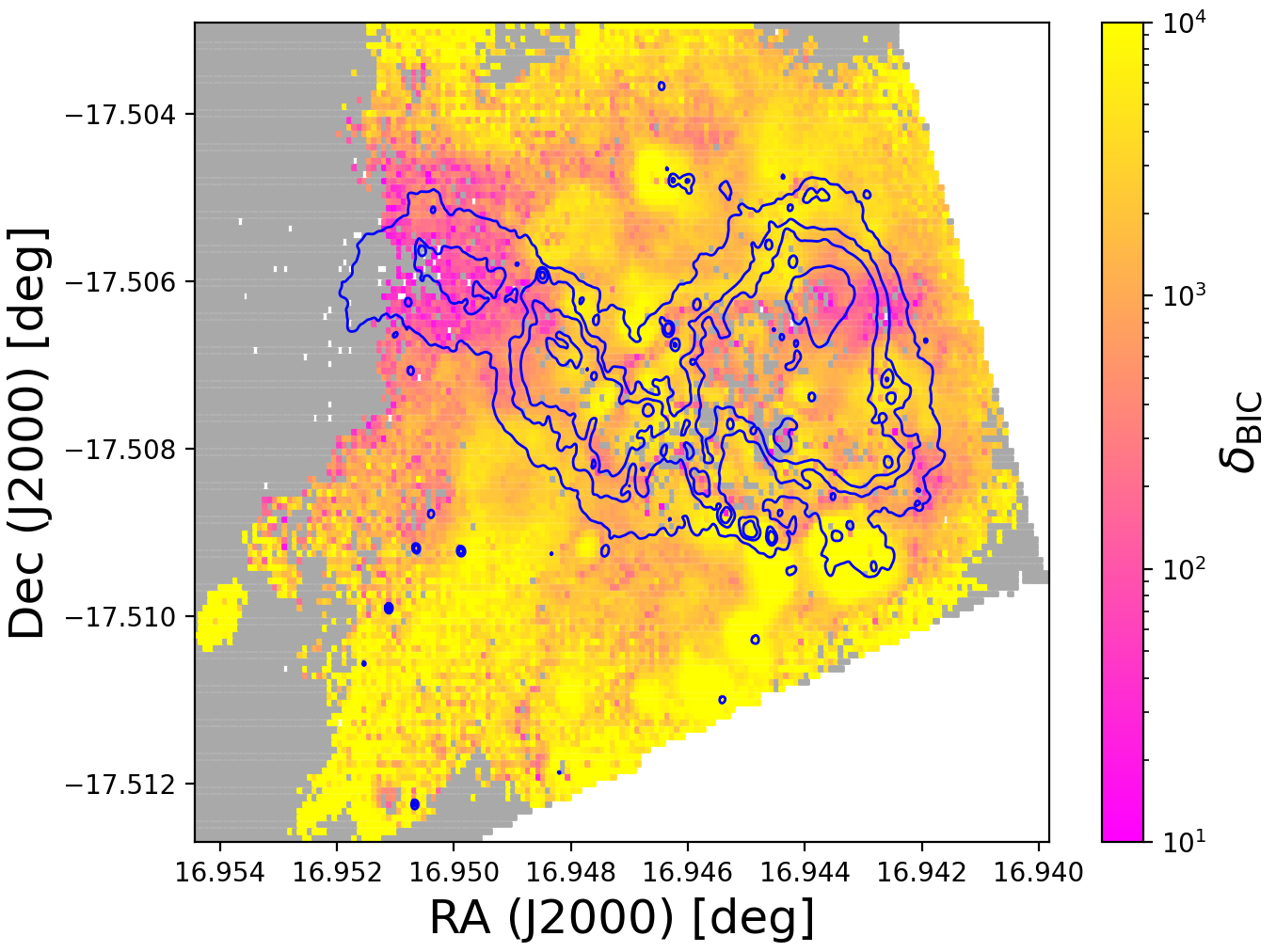}
    \caption{Spatial distribution of the $\delta_{\rm{BIC}}$ parameter measured from fits to the H$\alpha$ emission line. Spaxels with SNR$_{\rm{H\alpha}} < 10$ are masked (grey) and areas falling outside the combined IFU field of view are white. The blue contours are isointensity levels from the \emph{HST} ACS F814W image, as in Figure \ref{Figure_RGB}. The image demonstrates that a large fraction of spaxels contain emission line profiles that are best described with multiple kinematic components.}
    \label{Figure_BICMap}
\end{figure}

Many of the spectra in the KCWI and MUSE cubes required more than a single Gaussian component to accurately characterize the emission line profiles. In particular, 44\% of all spaxels had $\delta_{\rm{BIC}} \geq 1000$, 32\% had $\delta_{\rm{BIC}} \geq 2000$, and 6\% had $\delta_{\rm{BIC}} \geq 10000$. A map of $\delta_{\rm{BIC}}$ calculated from fits to the H$\alpha$ emission line in the MUSE cube can be seen in Figure \ref{Figure_BICMap}. Emission lines with very large $\delta_{\rm{BIC}}$ values tend to be concentrated in the various tidal arms to the south of the merger, in extended nebulae to the north, and in the overlap region. The NUV-bright tidal stream to the northeast (see Figure \ref{Figure_Atlas}) and across large areas of the western progenitor have $\delta_{\rm{BIC}}$ values consistent with single-component emission line profiles.

\begin{figure}[h!]
    \centering    \includegraphics[width=0.45\textwidth]{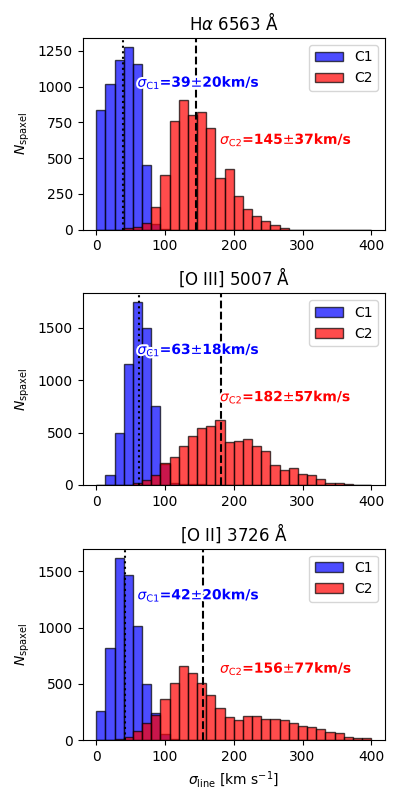}
    \caption{Velocity dispersion distributions for the two kinematic components (C1 and C2) fitted to the H$\alpha$, \oiii, and \oii~emission lines with $\delta_{\rm{BIC}} \geq 1000$ and SNR$_{\rm{line}} \geq 10$ across the entire merger. Dotted and dashed vertical lines mark the medians of the C1 and C2 distributions, respectively. The histograms demonstrate that the overwhelming majority of emission line profiles with $\delta_{\rm{BIC}} \geq 1000$~have components with quite distinct velocity dispersions.}
    \label{Figure_SigmaDistributions}
\end{figure}

Since many of the emission line profiles are best described using two Gaussian components that differ most obviously in their linewidths, ``C1'' is defined as the \emph{narrower} component and ``C2'' as the \emph{broader} component for a given emission line fit. Observable differences in the line component widths likely result from a combination of microscopic and macroscopic velocity broadening, hence, throughout the paper the terms linewidth and velocity dispersion are used interchangeably. The velocity dispersion distributions for C1 and C2 (corrected for instrumental broadening) are shown in Figure \ref{Figure_SigmaDistributions} for the H$\alpha$, \oiii, and \oii~lines, for spaxels with SNR$_{\rm{line}} \geq 10$ and $\delta_{\rm{BIC,H\alpha}} >$ 1000. The median and standard deviation of the C1 component is similar between the three emission lines, i.e., $\bar{\sigma}_{\rm{C1}}^{\rm{H\alpha}} = 39\pm20$ km s$^{-1}$, $\bar{\sigma}_{\rm{C1}}^{\rm{\oiii}} = 63\pm18$, and $\bar{\sigma}_{\rm{C1}}^{\rm{\oii}} = 42\pm20$ (1-$\sigma$ standard deviation).

The C2 linewidth distribution also has a similar median and standard deviation between the three lines: $\bar{\sigma}_{\rm{C2}}^{\rm{H\alpha}} = 145\pm37$ km s$^{-1}$, $\bar{\sigma}_{\rm{C2}}^{\rm{\oiii}} = 182\pm57$, and $\bar{\sigma}_{\rm{C2}}^{\rm{\oii}} = 182\pm57$~for H$\alpha$, \oiii, and \oii. Based on these descriptive statistics and the histograms in Figure \ref{Figure_SigmaDistributions}, it is clear that C2 can be easily distinguished from C1 on the basis of the component velocity dispersion.

As can be seen in Figure \ref{Figure_SigmaDistributions}, there is a small overlap in the histograms, indicating a small number of spaxels ($\sim$5 \%) containing two line components with similar linewidths ($\Delta\sigma_{\rm{C2-C1}} \lesssim 50$ km s$^{-1}$.) Spatially, spaxels with low $\Delta\sigma_{\rm{C2-C1}}$ appear concentrated along the periphery of the tidal/disturbed arms of the western component that are populated with star clusters. The largest differential linewidths are seen in the dusty regions to the SE and N of the eastern progenitor. Regions around resolved young star clusters are characterized by a C2 component that is $\sim$200 km s$^{-1}$ broader than C1. Application of velocity dispersion cuts to the two distributions, e.g., $\sigma_{\rm{C1}} < 100$~and $\sigma_{\rm{C2}} \geq 100$~km s$^{-1}$, were not made since they did not significantly alter the results (only a small number of measurements are excluded for any sensible dispersion cut.)

\subsection{Velocity Dispersion Maps}
\label{sec:R2_Kinematics}

\begin{figure*}[ht!]
    \centering    \includegraphics[width=\textwidth]{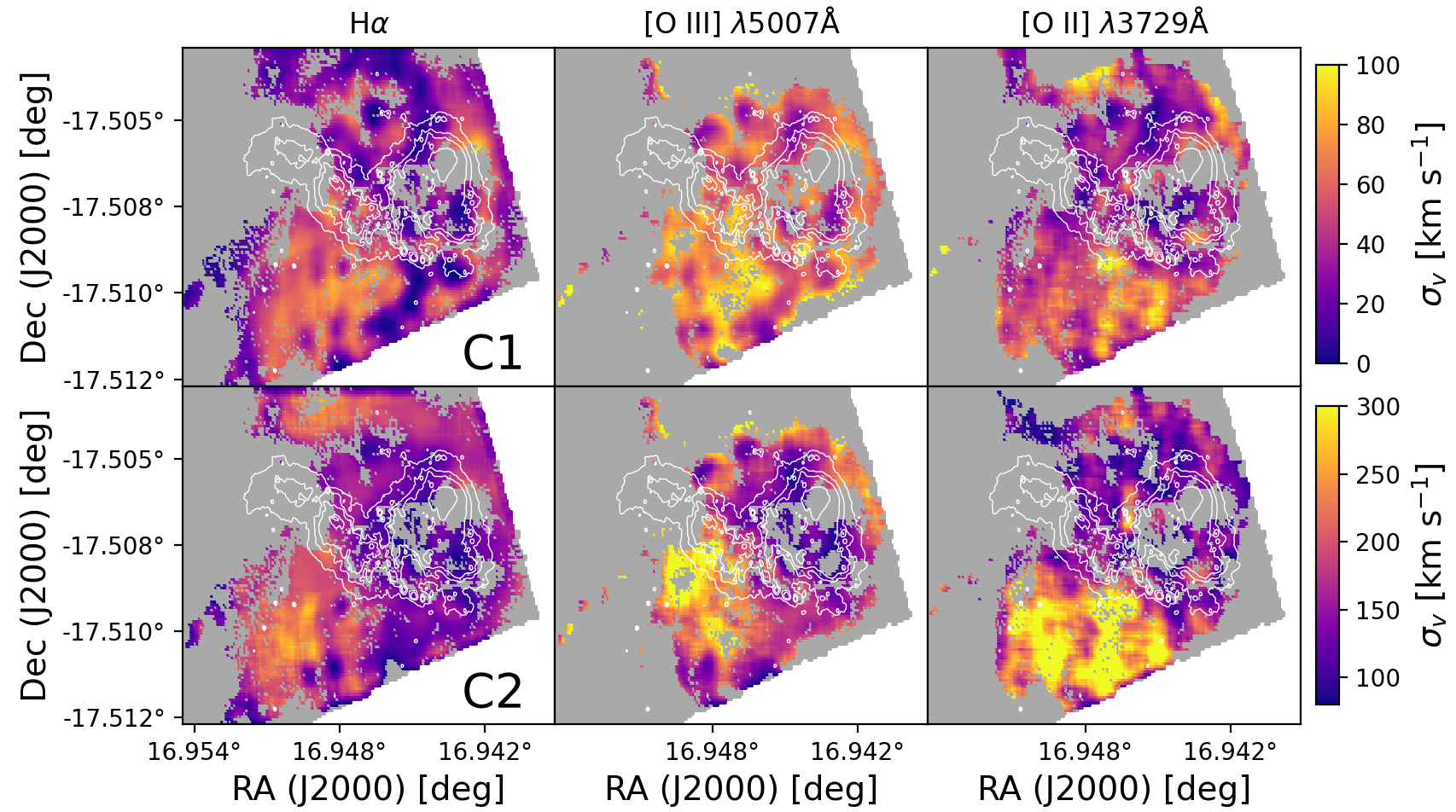}
    \caption{Velocity dispersion maps for the narrow (C1, top) and broad (C2, bottom) kinematic components of the H$\alpha$ (left), \oiii~$\lambda$5007 \AA~(middle), and \oii~$\lambda$3726 \AA~(right) emission lines with $\delta_{\rm{BIC}} \geq 1000$ and SNR$_{\rm{line}} \geq 10$. The top and bottom rows have colorbars scaled to the velocity dispersions spanned by C1 and C2 to highlight differences in their velocity structure. White contours are the same as in Figure \ref{Figure_RGB}. 
    }
    \label{Figure_DispersionMaps}
\end{figure*}

Velocity dispersion maps for the C1 and C2 components of the H$\alpha$, \oiii, and \oii~emission lines observed with MUSE and KCWI are presented in Figure \ref{Figure_DispersionMaps}. Maps include spaxels with $\delta_{\rm{BIC}} \geq 1000$ and SNR$_{\rm{line}} \geq 10$.
The colorbars are scaled according to the velocity dispersions spanned individually by C1 and C2 in order to highlight differences in their velocity fields. As a result of SNR cuts on each emission line, the higher-ionization \oiii~dispersion maps cover a smaller area than the recombination line and the low-ionization \oii~line, since H$\alpha$ has higher emissivity across a wider range of environments, and since O$^+$~can be formed at larger separations from ionizing sources than O$^{++}$. Masked regions are colored grey and the \emph{HST}/ACS F814W contours are shown in white for reference.

The C1 component of all three emission lines has large velocity dispersions ($\gtrsim$60 km s$^{-1}$) to the south, between the SE and SW tidal arms, in the overlap region, and to the north and west of VV\,114W. For reference, the various regions of VV 114 are indicated in the atlas shown in Figure \ref{Figure_Atlas}. All three emission lines show low C1 velocity dispersions across large regions of the star-forming disk of VV\,114W. While maps demonstrate that the \hii~regions along the SW tidal arm are associated with low C1 velocity dispersion in the recombination line and in \oiii, the C1 dispersions of the \oii~line are large in this region, suggesting the \oii-emitting gas is at a different physical depth along the line-of-sight, occupying a lower-ionization shell farther from the star clusters in the SE tidal arm. More generally, toward ionizing sources, the \oii- and \oiii-emitting gas may lie at different depths along the line-of-sight in a similar manner to what would be expected for the case of an idealized a Stromgen sphere with a stratified ionization structure. Therefore, for sight lines toward ionizing sources, the H$\alpha$ and \oiii~dispersion maps may probe different vertical ``layers'' of the ISM than the \oii~maps.

The C2 maps have, by definition, larger velocity dispersions than the C1 maps at every spaxel position. Spatially, the gas traced by the C2 component has a distinct velocity dispersion structure compared to C1, which is most obvious in the oxygen maps where C2 dispersions appear to vary more dramatically as a function of position. In particular, the \oiii~map shows a concentration of very high dispersion gas toward and to the south of VV 114E, and between the southern tidal arms. In the broad line \oii~map, there is a much larger contrast in velocity dispersion between VV\,114W and the regions to the south of the merger than is seen in the C1 \oii~map. Whereas the C1 and C2 maps both show smaller velocity dispersions toward the star-forming disk of VV\,114W, regions of high $\sigma_{\rm{C1}}$ are not generally regions of high $\sigma_{\rm{C2}}$, suggesting that the C2 component is emitted by gas that is physically distinct from the C1-emitting gas, and possibly responding to different physical drivers of turbulence.

\subsection{Optical Excitation Diagnostics}
\label{sec:R3_OpticalBPT}

\begin{figure*}[ht!]
    \centering
    \includegraphics[width=\textwidth]{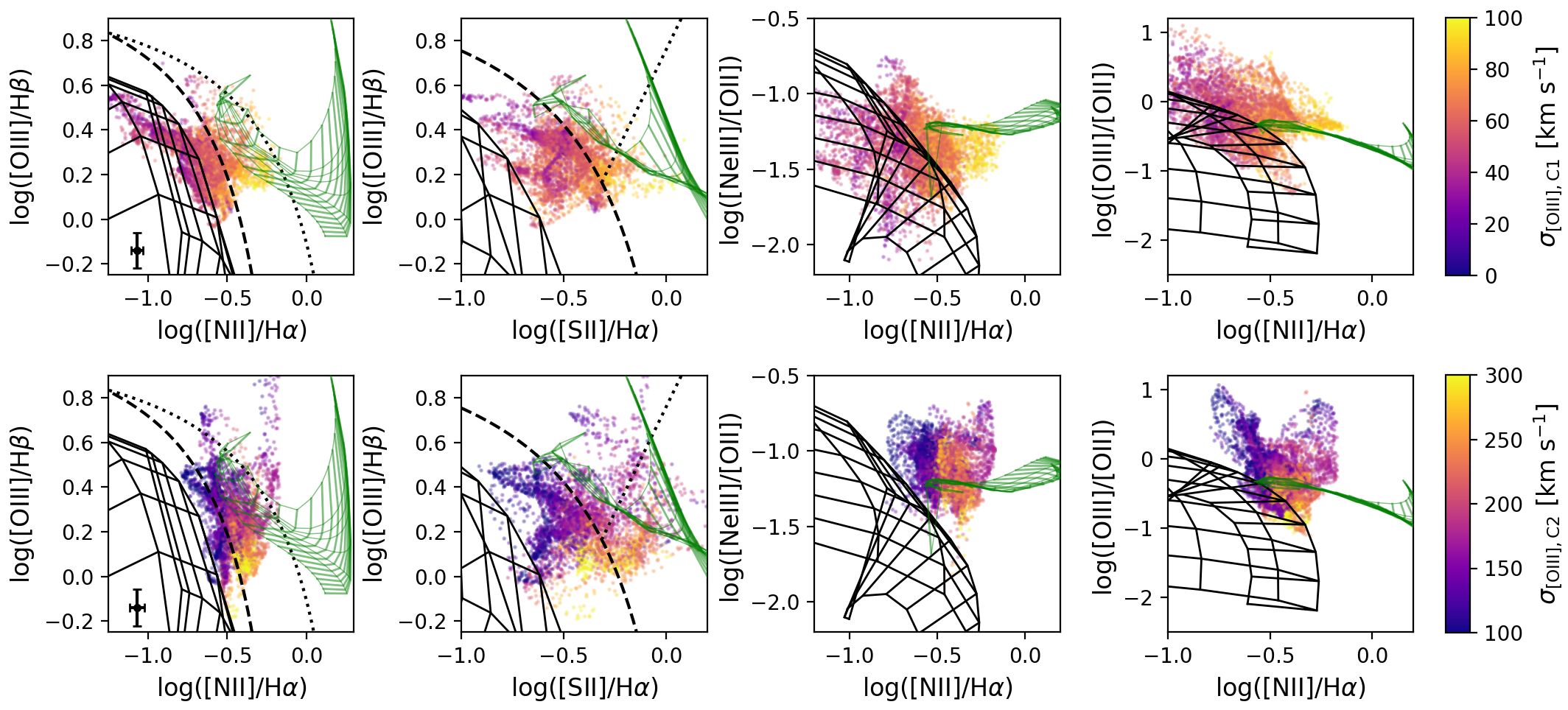}
    \caption{Optical diagnostic emission line ratios with fluxes from the C1 (top) and C2 (bottom) kinematic components using spaxels with SNR$_{\rm{line}} \geq 10$ and $\delta_{\rm{BIC}} \geq 1000$. Points are colored according to velocity dispersion in the \oiii~line. The black dashed and dotted lines in the \nii/\ha~diagrams (first column) are the \cite{Kauffmann03} empirical star-forming galaxy and the \cite{Kewley01} maximum starburst lines. The single filled black circles (placed at the bottom-left of the panels) have errorbars representing the median RMS uncertainties in the derived line ratios. The black dashed and dotted lines on the \sii/\ha~diagrams (second column) are the \cite{Kewley01} star-forming galaxy line and the \cite{Kewley06} LINER/Sy2 line. Black grids represent line ratios predicted by photoionization models spanning metallicities log(O/H)+12 from 7.6 to 9.2 and ionization parameters log($Q(H)$) from 6.5 to 8.5. The shock ionization models are shown as green grids that span a range in magnetic field strength $B$ from 10$^{-4}$ to 10 $\mu G$~and shock velocities $v_{\rm{shock}}$ from 100 to 500 km s$^{-1}$. The C1 line ratios are more often consistent with \hii~grids, whereas the C2 ratios more often occupy the regions between the \hii~and shock grids and within the shock grids themselves.}
    \label{Figure_BPTDiagrams}
\end{figure*}

Optical diagnostic line ratios from the MUSE and KCWI observations are shown in Figure \ref{Figure_BPTDiagrams} for spaxels with $\delta_{\rm{BIC}} \geq 1000$ and SNR$_{\rm{line}} \geq 10$. From left to right are the Baldwin-Phillips-Terlevich (BPT; \citealt{bpt81}), Veilleux-Ostriker (VO87; \citealt{Veilleux87}), \neiii/\oii~vs. \nii/H$\alpha$, and \oiii/\oii~vs. \nii/H$\alpha$~diagrams, with top and bottom rows showing the C1 and C2 components.
In the top and bottom rows, points are colored by the velocity dispersion of the C1 and C2 components of \oiii, respectively, with colorbars indicating the ranges. In the bottom-left corner of the \nii/\ha~panels (first column) are arbitrarily placed points with errorbars representing the median RMS uncertainty in the logarithmic line ratios, propagated from the flux density measurement errors in the datacubes.

Each diagram in Figure \ref{Figure_BPTDiagrams} features grids of photoionization (black) and radiative shock ionization (green) models composed from diagnostic line ratios predicted for a range of ionizing source, shockwave, and nebular properties. The grids were generated with the \textsc{MAPPINGS V}\footnote{https://mappings.readthedocs.io/en/latest/index.html} photoionization and shock modeling code \citep{dopitasutherland96,sutherlanddopita17}. The photoionization grids were constructed from a series \hii~region models. The ionizing spectral energy distribution (SED) was simulated with the \textsc{starburst99} spectral synthesis code \citep{leitherer99}, using a range of ionization parameters (6.5 $\leq$ log($Q$) $\leq$ 8.5). Model nebular spectra were generated by exposing model gas clouds with abundances varying between 7.6 $\leq$ log(O/H) + 12 $\leq$ 9.2 to each SED, for a total of 45 model \hii~regions (photoionization grid points). 

The \textsc{MAPPINGS V} shock model grids were acquired from the Mexican Million Models Shock Database (\textsc{3MDBs}\footnote{http://3mdb.astro.unam.mx:3686/}; \citealt{morisset14,alarie19}.) The shock model grid extracted from this database has the same input parameters as the \cite{allen08} \textsc{MAPPINGS III} library of fast radiative shocks, but was recalculated with \textsc{MAPPINGS V}. In particular, the shocked gas has solar metallicity, a preshock density of 1 cm$^{-1}$, while the transverse magnetic field and shock velocity varied between $10^{-4} \leq B \leq 10$~$\mu$G and 100 $\leq v_{\rm{shock}} \leq$ 500~km s$^{-1}$. The models predict line emissivities for gas downstream from the shock that is photoionized by the extreme UV and X-ray portions of the radiation field (mostly EUV photons for $v_{\rm{shock}} \lesssim 200$ km s$^{-1}$) emitted by hot gas cooling immediately behind shock fronts \citep{dopita03}. 

\begin{figure*}[ht!]
    \centering
    \includegraphics[width=\textwidth]{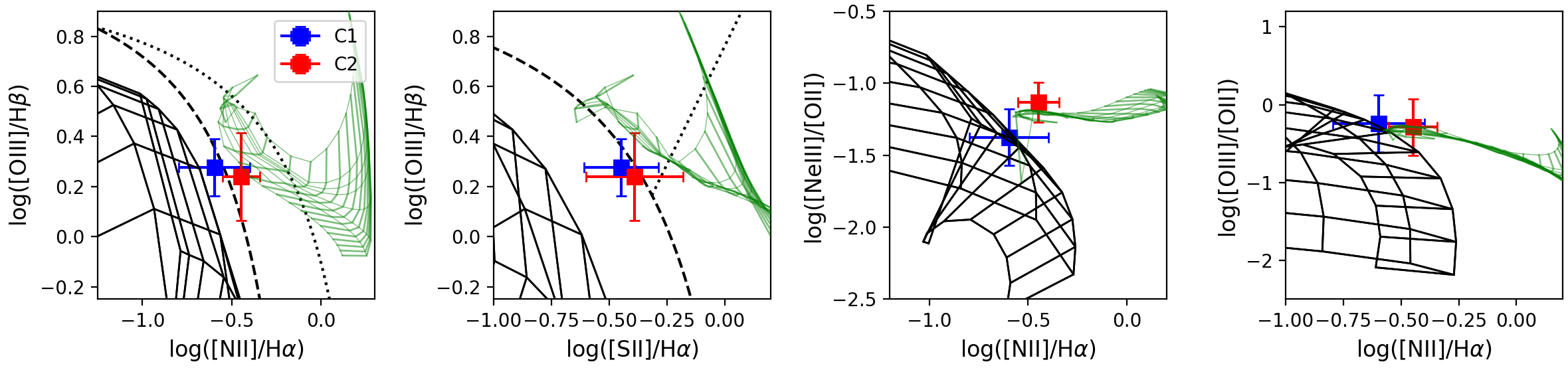}
    \caption{Diagnostic excitation diagrams as in Figure \ref{Figure_BPTDiagrams} but showing the median line ratios of C1 and C2 as blue and red boxes, with standard deviations of the distributions represented with errorbars. Lines and grids are the same as in Figure \ref{Figure_BPTDiagrams}.}
    \label{Figure_BPTMedians}
\end{figure*}

The excitation diagrams in Figure \ref{Figure_BPTDiagrams} demonstrate the C1 line ratios are more often consistent with the \hii~grids, whereas the C2 ratios more often occupy the region between the \hii~and shock grids and within the shock grids themselves. The (log) median \nii/H$\alpha$~values for the C1 and C2 distributions are -0.58$\pm$0.19 and -0.45$\pm$0.10, and for \oiii/H$\beta$, they are 0.27$\pm$0.11 and 0.23$\pm$0.17. The (log) median ratios of \sii/H$\alpha$, \neiii/\oii, and \oiii/\oii~are -0.44$\pm$0.15, -1.38$\pm$0.19, and -0.26$\pm$0.36 for C1, and -0.39$\pm$0.20, -1.15$\pm$0.14, and -0.31$\pm$0.35 for C2. The aggregate positions of the C1 and C2 measurements are shown in Figure \ref{Figure_BPTMedians}. Although the distributions have significant overlap, it is clear from Figure \ref{Figure_BPTMedians} that C2 is more consistent with shock ionization, with the largest separation in the \neiii/\oii~vs. \nii/H$\alpha$~diagram. There is an almost vertical continuation of points seen in the BPT diagrams, more clearly in C2. These points are spatially concentrated towards the \hii~regions of the SW tidal arm. The C2 lines also show a distinct sequence of large \oiii/\oii~ but small \nii/H$\alpha$~ratios (with small velocity dispersions), all occurring on the western edge of the VV\,114W disk and toward the \hii~regions in two southern tidal arms. This last sequence may be the result of \oiii~and \oii~probing gas at different physical depths, giving unphysically large \oiii/\oii~ratios in sightlines passing through regions with a high degree of ionization stratification.

For both C1 and C2, there is a clear gradient in velocity dispersion versus position on the diagrams, with $\sigma_{\rm{[OIII]}}$ increasing from left to right (moving from the \hii~grids to the shock grids). Additionally, the C2 distributions show a gradient in direction perpendicular to the \hii-shock sequence such that the regions with the largest C2 velocity dispersion have elevated \nii/H$\alpha$ and \sii/H$\alpha$ but relatively low \oiii/H$\beta$, \neiii/\oii, and \oiii/\oii.

\subsection{Theoretical Shock Mixing Sequence}
\label{sec:R4_MixingSequence}
\begin{figure*}[ht!]
    \centering
    \includegraphics[width=\textwidth]{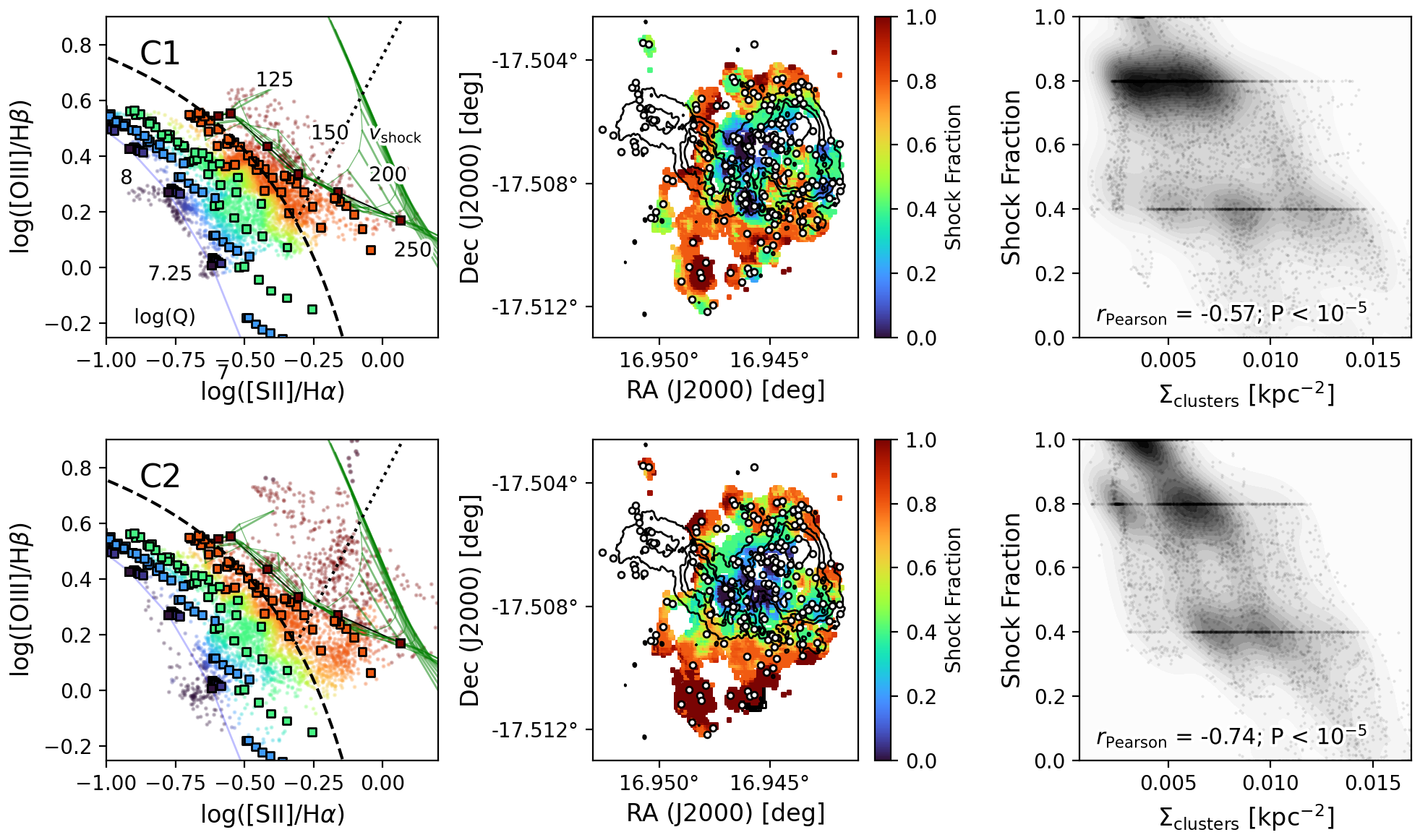}
    \caption{Left panels: BPT diagrams with the theoretical photoionization-shock mixing sequence represented by filled squares colored by shock fraction: dark blue corresponds to line ratios predicted from 0\% shocks (100\% photoionization), and dark red is 100\% shocks. Measured flux ratios are shown as small circles and are colored according to their position relative to the mixing sequence on the diagram. Middle panels: maps of the galaxy color-coded based on where the spaxel falls on the mixing sequence. The white filled circles are star clusters cataloged in \cite{linden21}, and contours are the same as Figure \ref{Figure_RGB}. Right panels: The grey filled contours are the KDE-smoothed distribution of shock fraction vs. local surface number density of star clusters for all spaxels (grey points). The analysis is repeated for the C1 components (top row) and C2 components (bottom row).}
    \label{Figure_MixingSequence}
\end{figure*}

The left panels of Figure \ref{Figure_MixingSequence} show the \oiii/\hb~vs. \sii/\ha~diagnostic diagrams for the C1 and C2 components (small points), overlaid with a theoretical shock mixing sequence (filled squares.) The mixing sequence is bracketed at low \sii/H$\alpha$ by the locus of \hii~region models with solar metallicity, shown as dark blue squares connected by a blue line, spanning ionization parameters from 6.5 $\leq$ log($Q$(H)) $\leq$ 8.5, and at large \sii/H$\alpha$ by the sequence of radiative shock models with solar metallicity, magnetic field $B = 1~\mu$G, pre-shock density of 1 cm$^{-3}$, and spanning shock velocities 100 $\leq$ $v_{\rm{shock}}$ $\leq$ 250 km s$^{-1}$. A total of 54 unique log($Q$)-$v_{\rm{shock}}$ pairs result from the 9 \hii~models and 6 shock models. For each pair ($Q_i$, $v_{\rm{shock,j}}$), the \sii/H$\alpha$~and \oiii/H$\beta$~emission line ratios predicted for gas ionized by some mixture of stellar photoionization and shock ionization is given by 

\begin{equation}
    R_{\rm{mix}} = (1-f_{\rm{shock}})R_{\rm{\hii}}(Q_i) + f_{\rm{shock}}R_{\rm{shock}}(v_{\rm{shock,j}}),
\end{equation}

where $R_{\rm{\hii}}(Q,i)$ represents the predicted line ratio for the \hii~photoionization model with solar metallicity and the $i^{\rm{th}}$ log($Q$) value and $R_{\rm{shock}}(v_{\rm{shock,j}})$ represents the predicted line ratio for the shock model with solar metallicity, magnetic field $B = 1~\mu$G, preshock gas density $n = 1$~cm$^{-3}$, and the $j^{\rm{th}}$ $v_{\rm{shock}}$ value. $f_{\rm{shock}}$ is the shock fraction that defines the mixing sequence and takes on values 0, 0.05, 0.2, 0.4, 0.8, and 1. For a mixing fraction of unity, the line ratios are just those predicted by the shock models. Observed points in the leftmost panels of Figure \ref{Figure_MixingSequence} are colored based on the interpolated shock fractions $f$ of the mixing sequence points.

\section{Discussion}
\label{sec:Discussion}

\begin{figure}[ht!]
    \centering
    \includegraphics[width=0.45\textwidth]{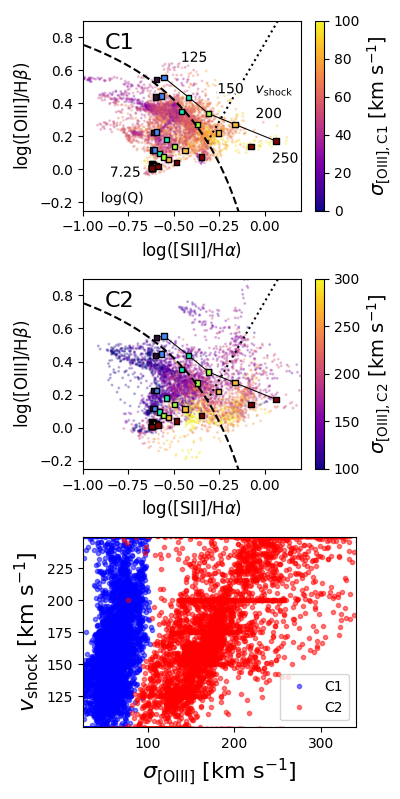}
    \caption{Top panels are BPT diagrams for the spaxels with $\delta_{\rm{BIC}} \geq 1000$~and SNR$_{\rm{line}} \geq 10$, colored by velocity dispersion in the \oiii~line. A mixing sequence is overlaid which contains line ratio predictions (filled boxes) for shock fractions of 0, 5, 20, 40, 80, and 100\%, bracketed on the low \sii/H$\alpha$ side by a single photoionization model with log($Q$) = 7.25 and on the high \sii/H$\alpha$~side by a shock model sequence with $B$ = 1 $\mu$G and spanning shock velocities between 100 and 250 km s$^{-1}$. The boxes are colored by increasing shock velocity (from blue to red). For the C2 component, there is a clear gradient in $\sigma_{\rm{\oiii}}$ in the direction of increasing shock velocity, which is demonstrated in the bottom panel.}
    \label{Figure_ShockVelocities}
\end{figure}

\subsection{Physical Nature of the C1 and C2 Components}
\label{sec:D1}

Across most of VV 114, the recombination lines and collisionally-excited lines feature a secondary broad component that accounts for almost half the optical line emission of the galaxy. The results demonstrate that the two components have velocity dispersions that differ by $\Delta\sigma \sim 120$~km s$^{-1}$, and the excitation diagrams show that while C1 is mostly photoionized gas, C2 is by varying degrees ionized by the hard radiation fields produced in interstellar shocks. The physical interpretation of the two components, based on these results, is developed in this Section.   

\subsubsection{The Narrow Component}
\label{sec:D1_narrowcomponent}
The integrated H$\alpha$ flux of the galaxy, from summing the C1 and C2 components of H$\alpha$ in spaxels with $\delta_{\rm{BIC}} \geq 1000$, SNR$_{\rm{H\alpha}} \geq 10$, and corrected for internal extinction using the Balmer decrement from the MUSE observations of H$\alpha$ and H$\beta$, is $F_{\rm{H\alpha}}$ = 8$\times$10$^{-13}$ erg s$^{-1}$ cm$^{-2}$ ($L_{\rm{H\alpha}} = 7\times10^{41}$ erg s$^{-1}$.) The integrated flux of the H$\alpha$ C1 component constitutes 54 percent of the integrated H$\alpha$ flux. The ratio is similar for H$\beta$ (51 percent) and ranges between 45 -- 54 percent for the forbidden lines. Most of the C1 flux in H$\alpha$ and \oiii~originates in VV\,114W and the tidal arms at positions surrounding resolved star clusters, but the \oii~emission is less well correlated with the location of bright clusters. 

Measured from the \oii, \oiii, and H$\alpha$ lines, the typical C1 linewidth is $\bar{\sigma}$ = 48 km s$^{-1}$. Correcting for a thermal broadening of $\sigma_{\rm{therm}}$ $\sim$20 km s$^{-1}$ at $T_{\rm{e}} \sim$ 10$^4$ K in the H$\alpha$~line ($\sim$5 km s$^{-1}$ for \oiii), and having already corrected for instrumental broadening, the corrected mean dispersion is $\bar{\sigma}_{\rm{corr}}$ = $\sqrt{\bar{\sigma}_{\rm{obs}}^{2} - \sigma_{\rm{inst}}^{2} - \sigma_{\rm{therm}}^{2}}$~= 46 km s$^{-1}$. Similar linewidths are commonly observed in many young star forming regions (e.g., \citealt{melnick21}), with  a variety of theories to explain their width beyond the thermal broadening contribution, including the turbulent stirring effect within star-forming regions caused by the motions and winds from low-mass stars, gravitational broadening caused by the virial motions of individual gas clouds in the ISM or the  multiple unresolved expanding shells. 

Turbulence driven by the random motion of low-mass stars and their bow shocks through star-forming clouds, i.e., the ``cometary'' shocks model of interstellar turbulence \citep{tenoriotagle93} can be immediately ruled out, since the data show no clear evidence for a positive correlation between luminosity and velocity dispersion in the H$\alpha$ lines, which is a prediction of the model and often observed in giant \hii~region \citep{melnick79,terlevich81}.

Another possibility is that the velocity dispersion of the C1 component arises from gravitational broadening, wherein the line emission originates from photoionized surfaces of multiple unresolved gas clouds within the spaxel. If these clouds are dynamically coupled through mutual gravitational interactions, and their motions are virialized, then the contribution to the observed velocity dispersion would follow $\sigma_{\rm{vir}} = \sqrt{GM/R}$. A crude estimate of the enclosed gas mass within each spaxel follows from assuming the molecular component (which dominates the ISM mass), plus the stellar mass component, can be evenly divided among the individual spaxels, and yields $M_{\rm{encl}}$ $\sim$10$^7$ $M_{\odot}$ per spaxel. Using $R \sim 60$~pc from the projected physical size of the spaxels, the velocity dispersion contribution from internal virial motions would be $\sigma_{\rm{vir}} \sim 30$~km s$^{-1}$. The emission line velocity dispersions, corrected for thermal and instrumental broadening, are still in excess of this contribution, requiring an addition source of turbulence for the C1-emitting gas.

Integrating unresolved expanding shells along the line-of-sight could result in supersonic emission line profiles and was proposed by \cite{chukennicutt94} as an explanation for the supersonic linewidths they measured in the giant \hii~region 30 Doradus. In the analytic model of \cite{MunozTunon96}, one effect of integrating unresolved expanding shells along the line of sight (LOS) is an anticorrelation between linewidths and their own flux (inclined bands), which are seen for the C1 component in the KCWI and MUSE data. The analytical model first developed by \cite{Dyson79} and further elaborated by \cite{tenoriotagle96} predicts that intrinsically Gaussian supersonic profiles $v_f$~should only arise from a collection of wind-driven shells if the distribution of shell ages is strongly peaked, which implies an unrealistic star formation history. For any other shell age distribution, the predicted line shapes are always flat-topped, which is not observed here. However, \cite{tenoriotagle96} demonstrate that if the intrinsic supersonic (usually flat-topped) emission line profiles are narrower than $v_f \approx 2\sigma_G$, where $\sigma_G = \sqrt{\sigma_{\rm{inst}}^2 + \sigma_{\rm{therm}}^2}$, then the observed profile will be Gaussianized due to the convolution with $\sigma_G$. For the MUSE data, $\sigma_{\rm{inst}} \sim 43$~km s$^{-1}$, hence $\sigma_G \sim \sqrt{43^2 + 20^2} \sim 47$~km s$^{-1}$. Therefore, the integration of unresolved shells along the LOS can contribute to a velocity dispersion of up to $\sim$100 km s$^{-1}$~before clear departures from Gaussian line profiles are observed. Thus, a combination of gravitational broadening by the virialized motions of unresolved clouds in the beam, as well as \hii~region expansion driven by massive stars can be used to explain the turbulent linewidths of the C1 component.

The placement of the C1 line ratios relative to the model grids shown in Figure \ref{Figure_BPTDiagrams} does not indicate the narrow line gas follows a shock mixing sequence, instead, the line ratios are overall consistent with the \hii~region models or fall behind the \cite{Kewley01} star-forming galaxy line. For the C1 component, the line velocity dispersion has no clear correlation with line ratios (Figure \ref{Figure_BPTDiagrams}) or with shock velocity (bottom panel, Figure \ref{Figure_ShockVelocities}), indicating the kinematics C1-emitting gas are not responding to shocks.

Overall, the C1 line ratios shown in Figure \ref{Figure_BPTDiagrams} are consistent with \hii~region-like photoionization rather than by shocks. The most likely interpretation of C1 is that it traces the photoionized ISM, with turbulent velocities dominated by \hii~region expansion plus a smaller contribution from the unresolved virial motions of the parent molecular clouds within the beam. The ionization and excitation of the C1-emitting gas is dominated by stellar ionizing radiation fields rather than shocks.

\subsubsection{The Broad Component}
\label{sec:D1_broadcomponent}

The velocity dispersion of C2 is on average 120 km s$^{-1}$ greater than C1. The IFU data do not indicate that the linewidth of C2 is correlated with its own flux, so cometary stirring by low-mass stars in star-forming regions can not be used as the primary explanation for the C2 linewidths. Furthermore, as shown in Section \ref{sec:D1_narrowcomponent}, gravitational broadening can only account for $\sigma \sim$~30 km s$^{-1}$ and is therefore not the origin for the large C2 linewidths. Finally, it seems rather unlikely that integration of unresolved expanding shells along the LOS can contribute up to 200-300 km s$^{-1}$ without any signs of line splitting at the resolution of the IFU data presented here, and as discussed in \ref{sec:D1_narrowcomponent}, if shell expansion contributes more than $\sim$100 km s$^{-1}$ of broadening, then flat-topped emission lines should have been observed. Clearly, some other mechanism is needed in order to explain the large velocity dispersions and line ratios of the C2-emitting gas.

In all of the diagnostic diagrams, the C2 component occupies the space between the photionization and shock grids, and in some cases the ratios fall directly on the latter. Furthermore, there is a gradient in the C2 velocity dispersions across the various BPT planes, generally showing an increased dispersion in the direction of increasing \nii/H$\alpha$ (or \sii/H$\alpha$) and decreasing \oiii/H$\beta$. The shock velocity at each spaxel was estimated in a similar manner to $f_{\rm{shock}}$~(c.f. Section \ref{sec:R4_MixingSequence}), i.e., by comparing the position on the BPT diagram with an interpolated surface of $v_{\rm{shock}}$ values built from the discrete points that comprise the mixing sequence (Figure \ref{Figure_ShockVelocities}.) To capture the general shape of the $v_{\rm{shock}}$ surface, a mixing sequence that connects a single \hii~region model at solar metallicity  and with log($Q$) = 7.25 to the locus of shock grids at solar metallicity and $B = 1$~$\mu$G (see the top two panels of Figure \ref{Figure_ShockVelocities}). In the bottom panel of the same Figure, blue points are the predicted shock velocity for C1 plotted against the velocity dispersion. No correlation can be seen here, since regions with any value of $\sigma_{\rm{C1}}$ see practically the entire range in predicted $v_{\rm{shock}}$ values. However, the same panel clearly shows that C2 velocity dispersions and shock velocity are correlated, indicating that the kinematics of C2 are responding to shocks. Interstellar shocks naturally explain both the emission line ratios and turbulent broadening of the C2 emission line components, but what is the source of shocks in this galaxy?

The origin of shocks, or shock-like signatures in the ISM of galaxies is varied. Cloud-cloud collisions, expanding supernova blastwaves, massive stellar winds, \hii~region expansion into the ambient ISM, turbulent mixing layers, and jet-ISM interactions are all examples of systems where shocks arise \citep{slavin93,dopitasutherland96,allen08}. Certainly, in the case of VV 114, large-scale tidal motions of the merger could be contributing to the appearance of shocks via compression in cloud-cloud collisions or shear-induced turbulent mixing layers \citep{rich11,rich15,slavin93}, but it is conceivable that the combined winds and supernovae from the numerous and widely-distributed star clusters are an additional, or even the dominant, source of interstellar shocks.

In the middle column of Figure \ref{Figure_MixingSequence}, regions of VV 114 are color-coded by shock fraction, and the locations of resolved star clusters from \cite{linden21,linden23} are shown as white filled circles. The right panels of the same figure are plots of shock fraction versus local surface number density of clusters (grey points). For the C2 component there is a strong anticorrelation between cluster density and shock fraction, such that shocks are more prevalent in the regions between and beyond the clusters rather than cospatial with the clusters themselves. This is consistent with a collimated galactic-scale wind (e.g., \citealt{medling15}), and/or with shock fronts from the widely-distributed clusters having an amplified effect on the ISM when they intersect in the interstitial regions between dense star-forming regions.

\subsection{Feedback from Cluster Winds}
\label{sec:D2_SSCWinds}

The total luminosity in H$\alpha$ contributed by shocks was estimated by summing the luminosity in the broad line components with $\sigma_{\rm{C2}} \geq 100$ km s$^{-1}$ from spaxels with SNR$_{\rm{H\alpha}} \geq 10$~and $\delta_{\rm{BIC}} \geq 1000$. The extinction-corrected total C2 H$\alpha$ luminosity is $L_{\rm{H\alpha,C2}}$ =  3.1$\times$10$^{41}$ erg s$^{-1}$. The threshold values adopted for $\sigma_{\rm{C2}}$, SNR$_{\rm{H\alpha}}$, and $\delta_{\rm{BIC}}$ are stringent enough that the derived $L_{\rm{H\alpha,C2}}$ is likely a lower limit. Using Figure 11 in \cite{rich10} and assuming an average shock velocity of 160 km s$^{-1}$, the total shock luminosity is $L_{\rm{shock}}$ = $75\times L_{\rm{H\alpha,C2}}$ = 2.3$\times$10$^{43}$ erg s$^{-1}$. 

The electron density of the warm ionized gas was estimated with \textsc{pyneb}\footnote{https://pypi.org/project/pyneb/} \citep{luridiana13} using single-component fluxes of the [S II] doublet, and assuming an electron temperature of $T_{\rm{e}}$ = 10$^4$ K. Spaxels in the IFU data are grouped based on proximity to cataloged star clusters, such that each spaxel is assigned to its nearest cluster. For each cluster, a radial density profile is constructed using the densities measured from the spaxels within its assigned region, and using their distances from the cluster. The density profiles are fit with models of steady-state, spherically expanding adiabatic winds (\citealt{cc85}, hereafter CC85.) The model was originally applied to galactic-scale superwinds driven by overlapping supernova explosions in the central starburst \citep{heckman_armus_miley90}, but has also been successfully applied on the scale of super star clusters (e.g., \citealt{wunsch07,murray11}). The model is completely determined by the the total energy and mass deposition rates ($\dot{E}_{\rm{T}}$ and $\dot{M}_{\rm{T}}$), which define the terminal asymptotic velocity $v_{\infty,\rm{A}}$ = (2$\dot{E}_{\rm{T}}/\dot{M}_{\rm{T}}$)$^{1/2}$, assumed in all our fits to be 2000 km s$^{-1}$ based on standard models of cluster-driven winds (e.g., \citealt{stevens_hartwell03}, \citealt{wunsch11}.) The 1-D hydrodynamic equations for the wind fluid are

\begin{equation}
    \frac{1}{r^2}\frac{d}{dr}(\rho u r^2) = q
\end{equation}

\begin{equation}
    \rho u \frac{du}{dr} = -\frac{dP}{dr}-q u
\end{equation}

\begin{equation}
    \frac{1}{r^2}\frac{d}{dr}\Big[\rho u r^2 \Big(\frac{1}{2}u^2 + \frac{\gamma}{\gamma-1}\frac{P}{\rho}\Big)\Big] = Q,
\end{equation}
\vspace{5pt}

where $r$ is the radial coordinate and $P$, $\rho$, and $u$ are the pressure, gas density, and velocity, and $\gamma$ is the ratio of specific heats ($\gamma = 5/3)$. For $r < R_{\rm{sc}}$, the volume-averaged mass and energy injection efficiencies are $q$ = $\dot{M}_{\rm{T}}/V$, $Q$ = $\dot{E}_{\rm{T}}/V$, with $V = 4\pi R_{\rm{sc}}^3 /3$. For $r > R_{\rm{sc}}$, $q = Q = 0$. The solutions for the Mach number $\mathcal{M}$ are given in \cite{cc85} as 

\begin{equation}
    \Big[\frac{(\gamma-1)+2/\mathcal{M}^2}{\gamma+1}\Big]^{(1+\gamma)/[2(5\gamma+1)]}\Big[\frac{3\gamma+1/\mathcal{M}^2}{1+3\gamma}\Big]^{-(3\gamma+1)/(5\gamma+1)} = \frac{r}{R_{\rm{sc}}},
\end{equation}

for $r < R_{\rm{sc}}$ and 

\begin{equation}
    \mathcal{M}^{2/(\gamma - 1)}\Big(\frac{\gamma-1+2/\mathcal{M}^2}{1+\gamma}\Big)^{(\gamma+1)/(2(\gamma-1)} = \Big(\frac{r}{R_{\rm{sc}}}\Big)^2.
\end{equation}

for $r > R_{\rm{sc}}$. With dimensionless radius $r_* = r/R_{\rm{sc}}$, the dimensionless velocity and density $u_*$ and $\rho_*$, expressed as

\begin{equation}
    u = u_*\dot{M}_{\rm{T}}^{-1/2}\dot{E}_{\rm{T}}^{1/2},
\end{equation}

\begin{equation}
    \rho = \rho_*\dot{M}_{\rm{T}}^{3/2}\dot{E}_{\rm{T}}^{-1/2}R_{\rm{sc}}^{-2},
\end{equation}

can be solved as a function of radius using the radial Mach number profiles:

\begin{equation}
    u_* = \mathcal{M} \Big(\frac{\dot{E}_{\rm{T}}}{\dot{M}_{\rm{T}}}\Big)^{1/2}  \Big(\frac{(\gamma-1)\mathcal{M}^2 + 2}{2(\gamma-1)}\Big)^{-1/2},
\end{equation}

\begin{equation}
    \rho_* = \Big(\frac{\dot{M}_{\rm{T}}}{4\pi u_*}\Big) \frac{r_*}{R_{\rm{sc}}} \hspace{10pt} (r < R_{\rm{sc}})
\end{equation}

\begin{equation}
    \rho_* = \Big(\frac{\dot{M}_{\rm{T}}}{4\pi u_* (r_* R_{\rm{sc}})^2}\Big) \hspace{10pt} (r > R_{\rm{sc}}).
\end{equation}

Interior to the radius $R_{\rm{sc}}$, individual massive stars inject energy and mass in the form of hot, low density winds, which propagate and intersect with one another, creating an overpressurized zone with a large, constant electron temperature. Beyond $R_{\rm{sc}}$, the pressure and electron density fall off as $1/r^2$.  Although star clusters with ages greater than 50 Myr, corresponding to the lifetime of stars with $M = 8M_{\odot}$, no longer contribute type II SNe or massive stellar winds (except for slower asymptotic giant branch stellar winds), they are not excluded from our wind analysis since their winds may still be propagating through the ISM beyond $R_{\rm{sc}}$. Star clusters which show an increasing radial electron density profile, or where the density profile is too sparsely sampled, are excluded from the analysis, resulting in a final sample of 66 clusters out of the total 180 from the \cite{linden21} \emph{HST} catalog. In all wind model fits, the parameter $v_{\infty}$ is fixed at 2000 km s$^{-1}$, and $\dot{E}_{\rm{T}}$ and $R_{\rm{sc}}$ are free to vary independently between the limits $10^{38} \leq \dot{E}_{\rm{T}} \leq 5\times 10^{42}$ erg s$^{-1}$ and $10 \leq R_{\rm{sc}} \leq 2\times 10^3$ pc. Individual star clusters are expected to have injection radii of order several tens of pc \citep{silich04,wunsch07,wunsch11}, significantly below the angular resolution of the IFU observations presented here. The radial density profiles show turnover radii of order several hundred pc, which is likely attributable to the combined effect of multiple unresolved and/or embedded star clusters within the same region all contributing to a combined multi-cluster wind. 

\begin{table*}[ht]
\centering
\caption{Derived properties of individual star clusters in VV 114. The full table can be found in Appendix \ref{appendixC}.}
\label{Table1}
\begin{tabular}{llllllll}
\hline
\hline
ID$^{a}$ & R.A. {[}deg{]} & Dec {[}deg{]} & log($M/M_{\odot}$)$^{a}$ & log($\alpha$/yr)$^{a}$ & $R_{\rm{sc}}$ {[}kpc{]} & $\dot{E}_{\rm{T}}$ {[}10$^{41}$~erg s$^{-1}${]} \\
\hline
20     & 16.950256      & -17.505949     & 5.63328                & 8.0845               & 0.30                    & 0.48                                           \\
27     & 16.948508      & -17.506545     & 4.48672                & 6.01317              & 1.10                    & 4.64                                           \\
32     & 16.947912      & -17.50706      & 4.37472                & 7.23817              & 0.57                    & 2.38                                           \\
37     & 16.947986      & -17.507059     & 7.64905                & 8.28497              & 0.77                    & 3.90                                           \\
39     & 16.947811      & -17.507157     & 5.0789                 & 7.2159               & 0.54                    & 1.76                                           \\
40     & 16.947009      & -17.507432     & 5.03637                & 6.61453              & 0.40                    & 1.11                                           \\
42     & 16.947036      & -17.505462     & 6.1175                 & 6.97089              & 0.60                    & 15.51                                          \\
44     & 16.946733      & -17.507178     & 5.49096                & 8.28497              & 0.51                    & 0.80                                           \\
47     & 16.948495      & -17.50992      & 5.23679                & 6.70362              & 0.31                    & 8.84                                           \\
53     & 16.946749      & -17.5076       & 4.00205                & 6.79271              & 0.37                    & 0.93                                           \\
 
\hline
\multicolumn{4}{l}{\small {\bf{Note.}} $^a$ values taken from \cite{linden21}.} \\
\end{tabular}
\end{table*}

\begin{figure}[ht!]
    \centering
    \includegraphics[width=0.46\textwidth]{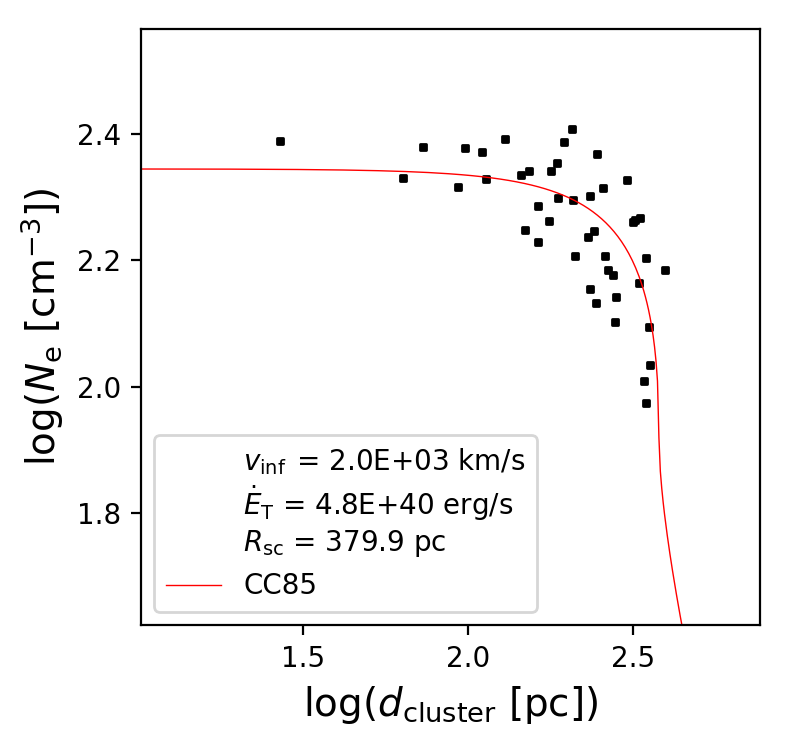}
    \caption{Example CC85 model fit to the radial density profile of ionized gas surrounding cluster \#84. The \sii-derived electron density in spaxels at increasing projected separation from the star-forming cluster are shown as black filled squares, incuding only spaxels closer to this cluster than to any other. The red line is the best-fitting CC85 cluster wind model to the data. The rest of the model fits are shown in Figure \ref{Figure_ClusterWindFits} in Appendix \ref{appendixC}.}
    \label{Figure_ClusterWindExample}
\end{figure}

An example fit of the adiabatic wind model to the radial electron density profile is shown in Figure \ref{Figure_ClusterWindExample} for cluster \#84. A subset of the analyzed clusters and their derived wind properties are listed in Table \ref{Table1}. The full sample is presented in Tables \ref{Table A1} and \ref{Table A2} in Appendix \ref{appendixC}. For approximately 40\% of the selected clusters, the radial density profiles are flat. It is possible that the characteristic turnover to inverse $r^2$ densities is beyond the probed radius, which is limited by proximity to the neighboring clusters. In these cases, $R_{\rm{sc}}$ is fit to the farthest radial point in the measured profile, making $R_{\rm{sc}}$ and $\dot{E}_{\rm{T}}$ \emph{lower limits} for such clusters, as indicated in Tables \ref{Table A1} and \ref{Table A2}. The ensemble of fitted $\dot{E}_{\rm{T}}$ and $R_{\rm{sc}}$ values (after using 1-$\sigma$ clipping to remove the 5 clusters with $\dot{E}_{\rm{T}} > 6.7\times 10^{41}$~ erg s$^{-1}$) shows a moderate correlation ($r_{\rm{Pearson}}$ = 0.64; $p$-value = 3$\times$10$^{-8}$), and similarly if only considering the models fit to flat profiles ($r_{\rm{Pearson}}$ = 0.74; $p$-value = 10$^{-4}$.) In light of the fact that $\dot{E}_{\rm{T}}$ and $R_{\rm{sc}}$ are free parameters, such a correlation must be driven by the data rather than a functional relationship between the model parameters. Thus, the cluster-driven wind model is a reasonable interpretation of the spatial density distribution in VV 114. By summing the fitted $\dot{E}_{\rm{T}}$ values across the full sample, the total energy injection rate from the resolved star clusters is $\Sigma \dot{E}_{\rm{T}}$ $\gtrsim$ 1.6$\times$10$^{43}$ erg s$^{-1}$, or 70\% $L_{\rm{shock}}$.

In the absence of AGN, the soft band X-ray (0.3-2.0 keV) flux provides an independent measure of the energy input from massive stellar winds and SNe. The diffuse thermal X-ray emission encompassing both progenitors has a total absorption-corrected soft band X-ray luminosity of $L_{\rm{X}}$ = 2$\times$10$^{41}$ erg s$^{-1}$ \citep{Grimes06}. With an efficiency of $\varepsilon_{\rm{X}}^{\rm{eff}}$ = 0.4--10 per cent, massive stellar winds and supernova explosions convert their mechanical energy into heating the ISM, forming a hot, X-ray luminous diffuse medium with emission peaking in the soft X-ray band \citep{OtiFloranes10,Franeck22}. The total soft-band X-ray luminosity is 10$^{-4} \times L_{\rm{bol}}$ \citep{Grimes06}, implying $\varepsilon_{\rm{X}}^{\rm{eff}}$ = 1 per cent, hence the total stellar wind and supernova mechanical energy injection rate is $\Sigma\dot{E}_*$ $\sim$ 2$\times$10$^{43}$ erg s$^{-1}$, in close agreement with the value obtained from fitting the CC85 cluster wind model to the density profiles around young clusters. 

Another point of comparison for the total cluster energy injection rate would be the global energy injection rate estimated from the star formation rate (SFR). The infrared bolometric luminosity of VV 114 is $L_{\rm{IR}}$ = 4.5$\times$10$^{11}$ $M_{\odot}$ \citep{sanders03}, corresponding to a SFR of 66 $M_{\odot}$ yr$^{-1}$. Since the star formation is evenly divided between the east and west components \citep{Goldader02}, the SFR for VV\,114W is 33 $M_{\odot}$ yr$^{-1}$. Taking this SFR as input, \texttt{starburst99} models \citep{leitherer99} imply a total kinetic energy injection rate from SNe and stellar winds in VV\,114W of $\Sigma\dot{E}_{\rm{K}}$ = 1.4$\times$10$^{43}$ erg s$^{-1}$, which is in very good agreement with the summed energy injection rate from the clusters.

The cluster wind analysis indicates that the combined stellar winds and SNe from the young massive star clusters distributed around the galaxy contribute at least $\sim$70\% of the turbulent energy in the ISM of the galaxy via shocks. Given the total mass in the merger, the binding energy of the molecular gas reservoir is $U = 3\times10^{58}$~erg and the dynamical time is $t_{\rm{dyn}}$ = 140 Myr. Following the analysis of NGC 3256 by \cite{rich11}, if all the shock energy is channeled into dissipating the total mechanical energy of the molecular gas, the dissipation timescale would be 40 Myr, meaning the gas should decouple and flow inward before rotational support can be restored. The dynamical time was computed assuming the molecular gas is supported against infall by rotation since $^{12}$CO(1-0) observations indicate a large N-S velocity gradient (e.g., \citealt{iono04}), however, several authors illustrate the difficulty in ascribing the complex molecular gas velocity field of VV 114 with pure rotation, instead suggesting the gas is already decoupling from the galactic potential and may be dominated by the radial or streaming motions of inflow (or outflow) \citep{Yun94,iono04}. Therefore, in the case of VV 114, shocks may already be acting as an efficient mechanism for funneling star-forming gas inward to fuel a future starburst, despite their tendency to locally disrupt giant molecular clouds on shorter timescales. The presence of shocks in the overlap region may explain ALMA observations of high dense gas fractions but low gas star formation efficiency there \citep{saito18}, by temporarily lowering the star formation efficiency locally but at the same time promoting the accumulation of dense gas as mechanical energy is being dissipated.

\subsection{Embedded Feedback}
\label{sec:D3_embeddedfeedback}

The maps of the \feii/\pab~near-IR emission line ratio derived from \emph{HST}/WFC3 narrow band imaging cover a large portion of VV 114, comparable to the maps of \nii/H$\alpha$ and \sii/H$\alpha$. The \nii, \sii, and H$\alpha$ fluxes discussed in this subsection were derived from single component fits to those emission lines in the MUSE data to facilitate comparison with near-IR line fluxes from the narrow band imaging data which do not distinguish multiple kinematic component fluxes. The signal of the narrow band \emph{HST} images was sufficient to map \feii/\pab~across most of the two progenitors and the overlap region, as well as the distorted spiral arm to the south. The two optical and one near-IR line ratio maps (smoothed using a Gaussian kernel with $\sigma$ = 1 pixel) have a similar large-scale pattern (see Figure \ref{Figure_NIRratiomaps} in Appendix \ref{appendixB}): regions with small line ratios are seen mostly in the western progenitor and elevated line ratios are seen in the eastern progenitor and overlap region. 

\begin{figure*}[ht!]
    \centering
    \includegraphics[width=\textwidth]{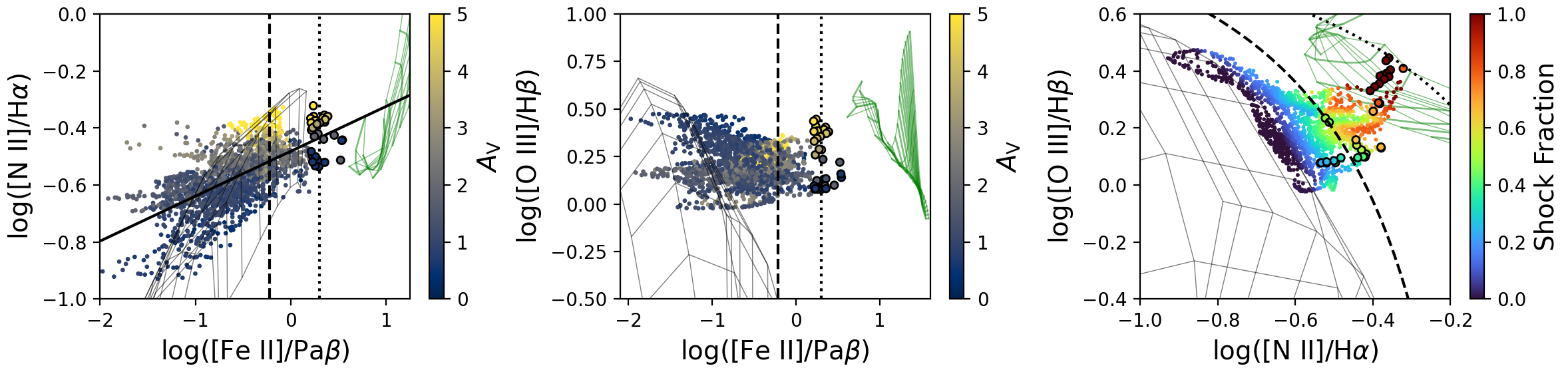}
    \caption{Comparisons between the optical and near-IR emission line ratios for areas of VV 114 covered by both MUSE and \emph{HST} are shown in the left and middle panels. The solid black line is a linear regression to the \nii/H$\alpha$~vs. \feii/\pab~data. The large circles outlined in black have log(\feii/\pab) $>$ 0.2, and are mostly located toward the highly extincted eastern component (regions enclosed in magenta boxes in the right panel of Figure \ref{Figure_NIRMapZoom}). Dashed and dotted vertical lines are empirically-established demarcations between H II, AGN, and shocked gas (e.g., \citealt{Riffel13}.) The right panel is a BPT diagram including spaxels inside the shared MUSE-\emph{HST} FoV, colored-coded by shock fraction, and showing the high \feii/\pab~regions as large black-outlined circles. Grey and green grids are photoionization and shock models.}
    \label{Figure_NIRBPT}
\end{figure*}

In Figure \ref{Figure_NIRBPT}, the spatial correlations between \nii/H$\alpha$, \oiii/H$\beta$, and the near-IR \feii/\pab~ratio are quantified. The left panel shows the distribution of \nii/H$\alpha$~vs. \feii/\pab~values across the entire distribution of pixels, showing a modest correlation ($r_{\rm{Pearson}} = 0.61$) within the shared MUSE-\emph{HST}~FoV. A linear fit to the distribution gives log(\nii/H$\alpha$) = 0.16~log(\feii/\pab) - 0.48. The lack of correlation for the \oiii/H$\beta$~vs. \feii/\pab~data is likely a result of  \oiii/H$\beta$~being much more sensitive to radiation field hardness than to shocks. Not shown is the \sii/H$\alpha$ vs. \feii/\pab~distribution, which also yields a modest correlation ($r_{\rm{Pearson}}$ = 0.65.) The correlations described here demonstrate that the near-IR \feii/\pab~emission line ratio is a good photoionization vs. shock indicator that also has the added benefit of probing conditions of gas embedded behind dust screens. 

Based on the linear fits to the \nii/H$\alpha$ vs. \feii/\pab~distribution, the lower and upper empirical near-IR demarcation lines \feii/\pab~= -0.22 and 0.3 \citep{Riffel13} correspond to log(\nii/H$\alpha$) = -0.52 and -0.43. While neither of these values are as large as the one-dimensional \cite{Kauffmann03} demarcation line at log(\nii/H$\alpha$) = -0.35 that efficiently separates starbursts from AGN and shocks in the BPT diagram (e.g., \citealt{Agostino23}), points above the upper demarcation line in \feii/\pab~would be clearly inconsistent with pure \hii-like photoionzation regardless of the \nii/H$\alpha$ ratio, which can be seen by comparing the dotted vertical line with the grey \hii~region grid in the left panel of Figure \ref{Figure_NIRBPT}. The figure also shows that \feii/\pab~is a more robust shock diagnostic since the \hii~and shock grids used here are more widely separated along that dimension ($\sim$1 dex) than for \nii/H$\alpha$ in the classical BPT diagram.

\begin{figure*}[ht!]
    \centering
    \includegraphics[width=\textwidth]{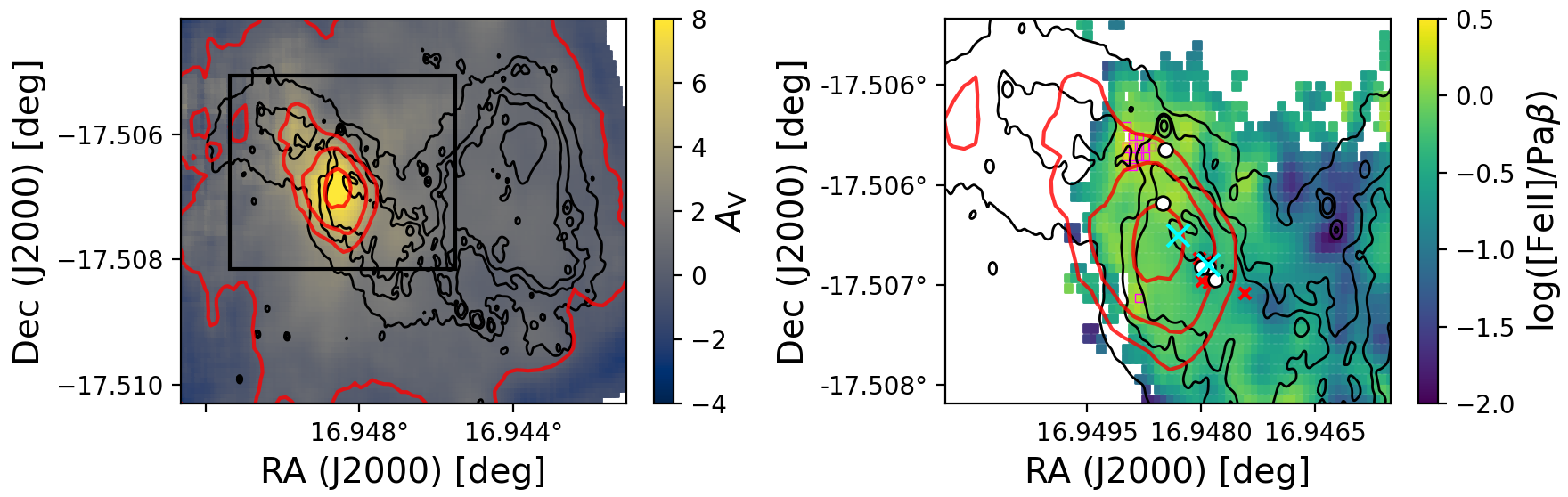}
    \caption{Left: Map of the visual extinction toward VV 114 calculated from the Balmer decrement and adopting the \cite{calzetti00} dust attenuation law ($R_V$ = 4.05). The black contours are the same as in Figure \ref{Figure_RGB}, the red contours are placed at values of $A_V$ = -2.5, 0, 3.5, 5, and 7.5 mag. The black box encloses the area displayed in the \feii/\pab~map in the right panel. The zoomed-in \feii/\pab~map reveals several regions of very high near-IR line ratios (outlined with magenta squares). Cyan crosses mark the locations of the two mid-IR-bright cores of VV 114E, and the white filled circles are star clusters located in the high extinction region.}
    \label{Figure_NIRMapZoom}
\end{figure*}

The eastern progenitor and overlap region, indicated with a black box in Figure \ref{Figure_NIRMapZoom}, contain most (24/34) of the pixels with log(\feii/\pab) $\gtrsim$ 0.2, the rest are located in the very bright northern star-forming region of VV\,114W (brightest F814W contour). These points are marked as large filled circles in Figure \ref{Figure_NIRBPT} and with open magenta squares in the right panel of Figure \ref{Figure_NIRMapZoom}, mostly in the high visual extinction ($A_V >$~6) regions to the north of the bright IR cores (cyan crosses). Although very elevated \nii/H$\alpha$~values are detected toward some of the same regions, the optical and near-IR emission lines are likely forming in physically different locations along the line-of-sight. In high extinction regions, caution must be taken when inferring excitation mechanisms from optical line ratios, since the detected optical emission may only originate on the near side of the dust screens, rather than the embedded region one may wish to probe. 

In the eastern nucleus, the near-IR and optical emission line ratios indicate varying degrees of shock excitation. Since this region has very high dust extinction, and since the near-IR observations have a higher optical depth, the natural conclusion is that in this region, the ionized gas is subject to feedback in the form of shocks, both at the near side of the dust layers and in the embedded ISM. There are several resolved star clusters in the vicinity of the nucleus of VV 114E (white filled circles on the map in Figure \ref{Figure_NIRMapZoom}), but they all have relatively low visual extinction, indicating that they are not substantially embedded. While these clusters may be contributing to the feedback seen in the optical emission in the region (e.g., cluster \#26 which is closest to the high \feii/\pab~region), the only obvious sources of feedback for the gas seen in the near-IR must be deeply embedded as well, possibly the embedded star-forming regions seen with VLA by \cite{song22} (red crosses in Figure \ref{Figure_NIRMapZoom}) or a nuclear starburst in the NW core, or even the AGN in the SW core, as revealed by \cite{Rich23} (who also detect strong shocks from elevated \feii/Pf$\alpha$ in this region).

\subsection{Galactic Superwind}

The shock fraction maps presented in Figure~\ref{Figure_MixingSequence}, especially for the C2 component, show a clear radially increasing gradient. Such large-scale structure is often observed in the [N\,II]/H$\alpha$ and [S\,II]/H$\alpha$ optical line ratios of objects with galactic superwinds (e.g., \citet{shopbell98,rich10,sharp10,westmoquette11,kreckel14}). The radial gradient may represent the true distribution of shocks, occurring preferentially away from the star-forming disk of VV\,114W, where interactions between the primary volume-filling wind fluid and the ambient medium are strongest. Bright (and narrow) line emission from the central starburst and dust extinction likely dilute the shock signal near the center of the galaxy, also contributing to the appearance of the radial gradient.

\begin{figure*}[ht!]
    \centering
    \includegraphics[width=\textwidth]{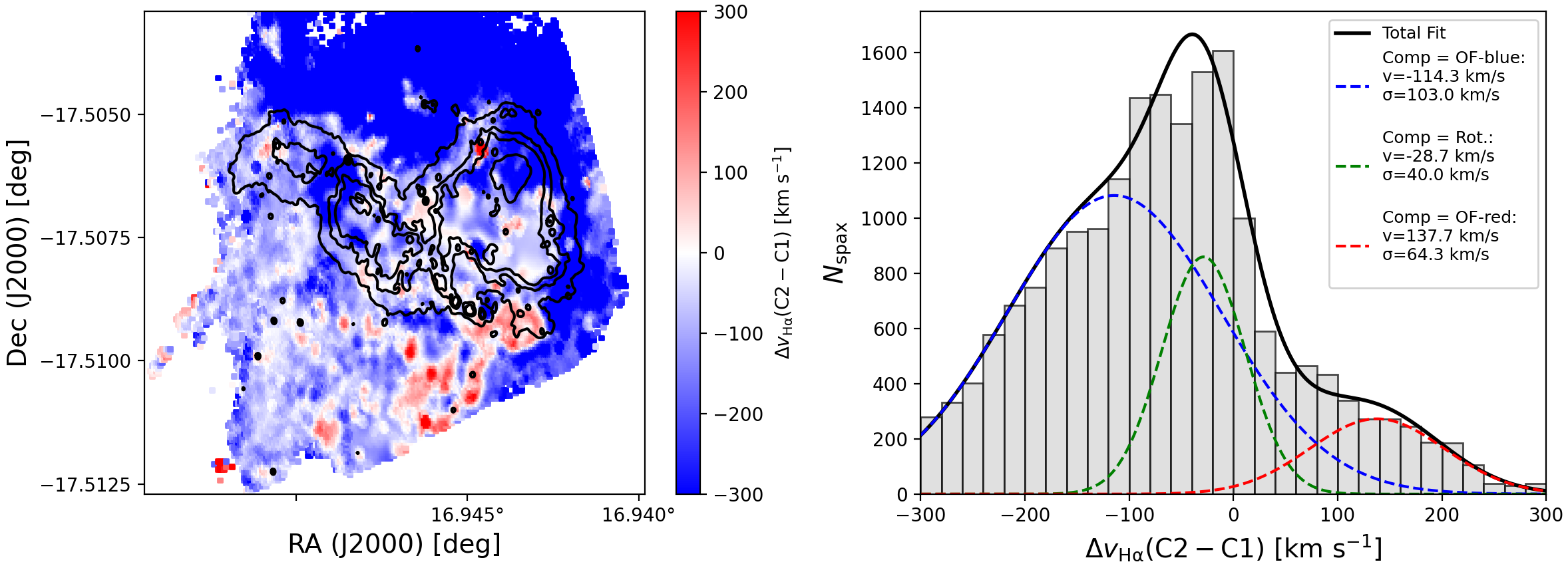}
    \caption{C2-C1 differential heliocentric radial velocity map (left) and distribution (right), derived from fits to the H$\alpha$ line in the MUSE datacube, the black contours are as in Figure \ref{Figure_RGB}. The best-fitting three-component Gaussian mixture model is overlaid on the histogram: the blue and red components represent regions of VV 114 where the broad C2-emitting gas blue- or redshifted velocities with respect to C1, the green component represents regions where the C2 component has similar velocities to C1 (rotation), and the black curve is the total model. The emission line velocities are consistent with a strong, broad emission line component that is blueshifted with respect to galactic rotation across much of the merger.}
    \label{Figure_veloffset}
\end{figure*}

Evidence for a superwind in VV\,114W was first reported by \citet{Grimes06}, who detected strong, broad interstellar absorption lines with a pronounced blueshifted component in the far-UV \emph{FUSE} spectra. The \emph{FUSE} observations imply outflowing material entrained by the wind with a projected velocity of 300-400 km s$^{-1}$. The narrow C1 component of H$\alpha$ likely traces large-scale rotation, showing a north-south velocity gradient that can also be clearly seen in the ALMA molecular gas velocities \citep{saito15}. As can be seen in the left panel of Figure \ref{Figure_veloffset}, the broad C2 component of H$\alpha$ is almost everywhere blueshifted by 100-300 km s$^{-1}$ with respect to the narrow C1 component, strongly suggesting non-rotational motions. This result was verified by producing moment 3 (skew) maps for several of the bright emission lines in the MUSE data, which show a close spatial correspondence with the differential velocity map, i.e., predominantly negative values of skew are found across the merger. The distribution of projected differential C2-C1 velocities ($\Delta v_{\rm{H\alpha}}$(C2-C1)) across the merger is shown in the right panel of Figure \ref{Figure_veloffset}, where it can be seen there are three distinct components: although some regions have C1 and C2 gas moving together (green dashed line), most of the broad component is blueshifted with respect to C1 (blue dashed line.) A third, redshifted component (red dashed line) can be seen in the right panel of Figure \ref{Figure_veloffset}: almost all of these positions are toward the SW tidal arm (highlighted in Figure \ref{Figure_Atlas}), but these kinematics likely reflect expansion of ionized bubbles or \hii~regions rather than the far-side of a galactic superwind, since at these positions the C2 and C1 components have similar, and relatively small, velocity dispersions and form double-peaked profiles (kurtosis of the emission lines tends to be large in these regions, also see Figures \ref{Figure_LineFits} and \ref{Figure_DispersionMaps}.)

\subsubsection{Dynamical State of the Wind}

\begin{figure}[ht!]
    \centering
    \includegraphics[width=0.45\textwidth]{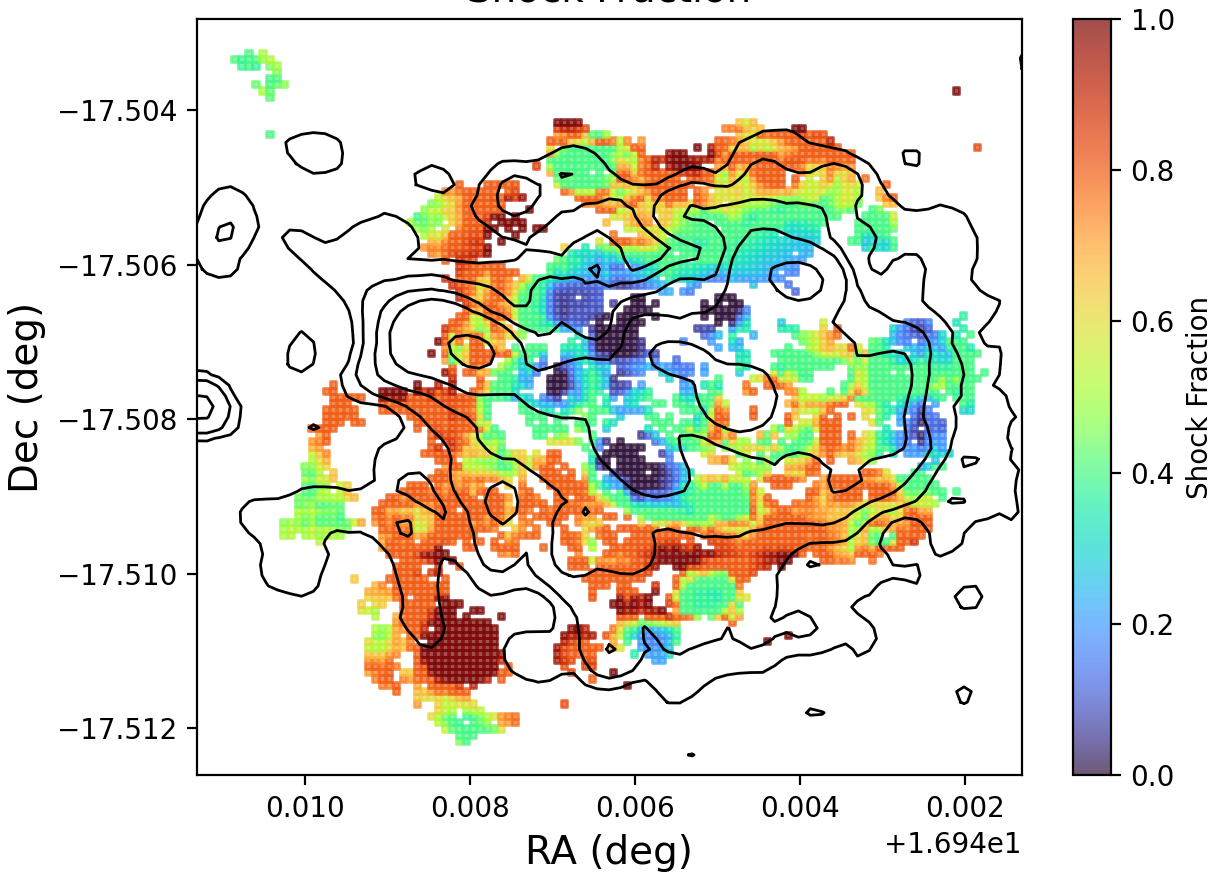}
    \caption{Spatially resolved shock fraction for the broad component (C2) ionized gas, traced by the optical emission line ratio \sii/H$\alpha$, as in Figure \ref{Figure_MixingSequence}. The \emph{Chandra} soft band (0.3-2.0 keV) X-ray image is represented by the black contours at the 4, 10, 20, and 80$\sigma$ levels. The resolved X-ray emission aligns closely with the spatial structure of shocks, suggesting they both trace the energetic interaction between a galactic wind fluid and the ambient ISM.}
    \label{Figure_xray_shock_comp}
\end{figure}

If the wind is isotropic ($4\pi$ sr opening angle) and is driven by the starburst, it should inflate a bubble of entrained ISM. Figure \ref{Figure_MixingSequence} clearly shows an abrupt transition in C2 shock fraction from $\sim$40 to 80\% that occurs at a radius of $r_{\rm{bubble}}$ $\sim$ 4 kpc (matching the $\sim$10$\sigma$ contour that encloses $\sim$70\% of the soft band X-ray flux in the \emph{Chandra} image, see Figure \ref{Figure_xray_shock_comp}), consistent with the shock-dominated wind-ISM interface defining the surface of a symmetric wind-blown bubble. In projection, this is coincident with the outer edge of VV 114W, determined by visual inspection of the \emph{HST} broad band F435W and F814W images. In the nomenclature of \cite{heiles90}, this is possibly consistent with the bubble undergoing blowout, where the outflowing material has reached the edge of the galactic disk (e.g., \citealt{maclow99}.)

The total kinetic energy in the wind-entrained ionized gas can be estimated using an assumed outflow velocity and by summing the mass of the C2-emitting gas, i.e., $M_{\rm{C2}}$ = $m_{\rm{p}}Q(\rm{H}^0)$/($n_{\rm{e}}\alpha_{\rm{B}}$), where $m_{\rm{p}}$ is the proton mass, $n_{\rm{e}}$ is the electron density, $\alpha_{\rm{B}}$ is the Case B recombination coefficient, and where $Q(\rm{H}^0)$, the number of ionizing Lyman continuum photons s$^{-1}$, can be computed from the extinction-corrected H$\alpha$ luminosity: $Q(\rm{H}^0)$ = ($L_{\rm{H\alpha}}$/$E_{\rm{H\alpha}}$) ($\alpha_{\rm{B}}$/$\alpha_{\rm{eff}}$) \citep{osterbrockferland06}. Assuming a deprojected wind velocity of $v_{\rm{exp}} = 2000$ km s$^{-1}$ (the same velocity used in the cluster wind models  $v_{\infty,\rm{A}}$), the estimated kinetic energy of the entrained ionized gas is $E_{\rm kin} \sim 8 \times 10^{58} n_{\rm{e}}^{-1} \text{ erg}$.

The energy budget of the wind can be assessed by comparing the observed kinetic energy injection rate, $\dot{E}_{\rm kin} = E_{\rm kin}/t_{\rm dyn}$, to the energy output from the starburst which was estimated in Section \ref{sec:D2_SSCWinds}. For $r_{\rm{bubble}} = 4$\,kpc and $v_{\rm exp} = 2000$\,km\,s$^{-1}$, the dynamical time (assuming a constant energy injection rate) is $t_{\rm{dyn}}$ = 2 Myr, yielding $\dot{E}_{\rm kin} \sim 1.2 \times 10^{45}$\,$n_{\rm{e}}^{-1}$\,erg\,s$^{-1}$. Adopting a reasonable value for the electron density in the outflowing ionized gas, $n_{\rm{e}}$ = 100 cm$^{-3}$ (close to the galaxy-wide average from the \sii~measurements), the total kinetic energy of the outflow is $E_{\rm kin} =$ 8 $\times$ 10$^{56}$ erg, and the kinetic energy injection rate is $\dot{E}_{\rm kin}$ = 1.2 $\times$ 10$^{43}$ erg s$^{-1}$, in very close agreement with the mechanical energy injection rate from the cluster winds that was estimated in Section \ref{sec:D2_SSCWinds} ($\Sigma\dot{E}_{\rm{T}} \sim 1.6 \times 10^{43}$\,erg\,s$^{-1}$), confirming that stellar winds and supernovae from the star clusters are the likely power source for the superwind.

\subsubsection{Galactic Impact of the Superwind}

The mass outflow rate ($\dot{M}_{\rm out}$) and mass loading factor ($\eta = \dot{M}_{\rm out}/\rm SFR$) are important for gauging the wind's impact on the galaxy. The mass outflow rate is $\dot{M}_{\rm out} = 10 M_{\rm{bubble,7}}V_{\rm{bubble,1000}}r_{\rm{bubble,kpc}}^{-1}$ $M_{\odot}~\rm{yr}^{-1}$, where $M_{\rm{bubble,7}}$ is the mass of ionized gas in 10$^7 M_{\odot}$ and $V_{\rm{bubble,1000}}$ is the expansion velocity in units of 1000 km s$^{-1}$. The total mass of outflowing ionized gas is $M_{\rm ion} = 10^7$\,$M_\odot$, and using $v_{\rm exp} = 2000$\,km\,s$^{-1}$, the mass outflow rate is $\dot{M}_{\rm out} \sim 5$\,$M_\odot$\,yr$^{-1}$, which is the same value derived by \cite{Grimes06}. The mass loading factor, assuming SFR = 33 $M_{\odot}$ yr$^{-1}$ in VV 114W, is $\eta \sim 0.2$. While in principle $M_{\rm{bubble,7}}$ includes both the mass of the entrained ionized medium and the primary wind fluid itself, the optical observations presented here do not provide a means of measuring the latter, therefore, the mass outflow rate and mass loading factor should be considered lower limits.

\section{Summary and Conclusions}
\label{sec:Conclusions}

\begin{enumerate}

\item Decomposition of optical emission line profiles from the wide-field IFU observations reveals a secondary, broad and blueshifted component across much of the galaxy, contributing nearly 50\% of the total H$\alpha$~flux.
\item The narrow component (C1) has line ratios consistent with stellar (\hii-region-like) photoionization, while the broad component (C2) is more consistent with a shock mixing sequence. C2 linewidths are correlated with shock velocities predicted from models of interstellar shocks.

\item The shock fraction in C2 gas is strongly anticorrelated with the surface number density of resolved star clusters, suggesting that, in addition to a possible tidal contribution to shocks from the merger itself, massive stellar winds a SNe contribute to shocks in the ISM.

\item The ISM around young massive clusters has electron density profiles consistent with predictions from adiabatic expanding cluster wind models. Models are fit to all young clusters then summed to give the predicted total starburst energy deposition rate which agrees well with the stellar mechanical energy input rate estimated from both the soft band X-ray luminosity and from the SFR. The resolved star clusters contribute the majority ($\sim$70\%) of the total shock energy in the merger.

\item The widespread shocks driven by winds from star clusters would be sufficient to cause the entire $\sim$10$^{10} M_{\odot}$~cold gas reservoir to dissipatively fall into the potential well of the merger in much less than a dynamical time, if the shock energy could be efficiently coupled with the cold gas mass.

\item Near-IR imaging data provide \feii/\pab~ratios across much of the merger. The near-IR ratio map has good spatial correspondence with the optical line ratio maps. Model grids of photoionization and shocks are better separated in optical-near-IR excitation diagrams involving the \feii/\pab~ratio than in the classical optical excitation diagrams.

\item Most of the regions with very high \feii/\pab~ratios occur north of the highly extincted nucleus of VV 114E, where optical ratios are also large. Since the area has large visual extinction, the \feii/\pab~ratio must be probing to higher optical depth, indicating feedback on both the exposed and veiled layers of VV 114E. The energy source for the embedded feedback in VV 114E is not clear, but might involve the nuclear starburst in the NW core or the AGN in the SW core.

\item The ionized gas kinematics, showing a blueshifted broad component across most of the merger, and the shock map together imply an ongoing galactic superwind which has entrained ambient material into a $\sim$4 kpc bubble that is possibly undergoing blowout. The superwind is powered by the starburst, since the energy budget of the bubble is in very close agreement with the kinetic energy input rate found from summing the individual cluster winds. The mass outflow rate is 5 $M_{\odot}$ yr$^{-1}$, in agreement with the value derived from far-UV spectroscopic observations \citep{Grimes06}.

\end{enumerate}

\begin{acknowledgments}
We thank the anonymous referee for their thoughtful feedback during the review process. The data presented herein were obtained at the W. M. Keck Observatory, which is operated as a scientific partnership among the California Institute of Technology, the University of California and the National Aeronautics and Space Administration. The Observatory was made possible by the generous financial support of the W. M. Keck Foundation. 
This research is based on observations made with the NASA/ESA Hubble Space Telescope obtained from the Space Telescope Science Institute, which is operated by the Association of Universities for Research in Astronomy, Inc., under NASA contract NAS 5–26555. These observations are associated with program HST-GO-17285.

Research at UC Irvine has been supported primarily by NSF AAG grant \#2408820. VU further acknowledges partial support from NASA ADSPS grant \#80NSSC25K7477. Support for program JWST-GO-01717 was provided by NASA through a grant from the Space Telescope Science Institute, which is operated by the Association of Universities for Research in Astronomy, Inc., under NASA contract NAS 5-03127.
TG acknowledges support from ARC Discovery Project DP210101945. MB thanks the financial support from the IAU-Gruber foundation fellowship. MSG acknowledges that this research project was
supported by the Hellenic Foundation for Research and Innovation (HFRI)
under the "2nd Call for HFRI Research Projects to support Faculty
Members \&
Researchers" (Project Number: 03382). AMM acknowledges support from the NASA Astrophysics Data Analysis Program (ADAP) grant number 80NSSC23K0750 and from NSF AAG grant \#2009416 and NSF CAREER grant \#2239807.

\end{acknowledgments}


\vspace{5mm}
\facilities{KeckII:KCWI, VLT:MUSE}

\bibliography{references}{}

\begin{thebibliography}{}
\expandafter\ifx\csname natexlab\endcsname\relax\def\natexlab#1{#1}\fi
\providecommand{\url}[1]{\href{#1}{#1}}
\providecommand{\dodoi}[1]{doi:~\href{http://doi.org/#1}{\nolinkurl{#1}}}
\providecommand{\doeprint}[1]{\href{http://ascl.net/#1}{\nolinkurl{http://ascl.net/#1}}}
\providecommand{\doarXiv}[1]{\href{https://arxiv.org/abs/#1}{\nolinkurl{https://arxiv.org/abs/#1}}}

\bibitem[{{Agostino} {et~al.}(2023){Agostino}, {Salim}, {Boquien}, {Janowiecki}, {Salas}, \& {Couto}}]{Agostino23}
{Agostino}, C.~J., {Salim}, S., {Boquien}, M., {et~al.} 2023, \mnras, 526, 4455, \dodoi{10.1093/mnras/stad3027}

\bibitem[{{Alarie} \& {Morisset}(2019)}]{alarie19}
{Alarie}, A., \& {Morisset}, C. 2019, \rmxaa, 55, 377, \dodoi{10.22201/ia.01851101p.2019.55.02.21}

\bibitem[{{Allen} {et~al.}(2008){Allen}, {Groves}, {Dopita}, {Sutherland}, \& {Kewley}}]{allen08}
{Allen}, M.~G., {Groves}, B.~A., {Dopita}, M.~A., {Sutherland}, R.~S., \& {Kewley}, L.~J. 2008, \apjs, 178, 20, \dodoi{10.1086/589652}

\bibitem[{{Armus} {et~al.}(2009){Armus}, {Mazzarella}, {Evans}, {Surace}, {Sanders}, {Iwasawa}, {Frayer}, {Howell}, {Chan}, {Petric}, {Vavilkin}, {Kim}, {Haan}, {Inami}, {Murphy}, {Appleton}, {Barnes}, {Bothun}, {Bridge}, {Charmandaris}, {Jensen}, {Kewley}, {Lord}, {Madore}, {Marshall}, {Melbourne}, {Rich}, {Satyapal}, {Schulz}, {Spoon}, {Sturm}, {U}, {Veilleux}, \& {Xu}}]{Armus09}
{Armus}, L., {Mazzarella}, J.~M., {Evans}, A.~S., {et~al.} 2009, \pasp, 121, 559, \dodoi{10.1086/600092}

\bibitem[{{Avery} {et~al.}(2021){Avery}, {Wuyts}, {F{\"o}rster Schreiber}, {Villforth}, {Bertemes}, {Chang}, {Hamer}, {Toshikawa}, \& {Zhang}}]{avery21}
{Avery}, C.~R., {Wuyts}, S., {F{\"o}rster Schreiber}, N.~M., {et~al.} 2021, \mnras, 503, 5134, \dodoi{10.1093/mnras/stab780}

\bibitem[{{Baldwin} {et~al.}(1981){Baldwin}, {Phillips}, \& {Terlevich}}]{bpt81}
{Baldwin}, J.~A., {Phillips}, M.~M., \& {Terlevich}, R. 1981, \pasp, 93, 5, \dodoi{10.1086/130766}

\bibitem[{{Bennett} {et~al.}(2013){Bennett}, {Larson}, {Weiland}, {Jarosik}, {Hinshaw}, {Odegard}, {Smith}, {Hill}, {Gold}, {Halpern}, {Komatsu}, {Nolta}, {Page}, {Spergel}, {Wollack}, {Dunkley}, {Kogut}, {Limon}, {Meyer}, {Tucker}, \& {Wright}}]{hinshaw13}
{Bennett}, C.~L., {Larson}, D., {Weiland}, J.~L., {et~al.} 2013, \apjs, 208, 20, \dodoi{10.1088/0067-0049/208/2/20}

\bibitem[{{Bertin} \& {Arnouts}(1996)}]{bertin96}
{Bertin}, E., \& {Arnouts}, S. 1996, \aaps, 117, 393, \dodoi{10.1051/aas:1996164}

\bibitem[{{Buiten} {et~al.}(2024){Buiten}, {van der Werf}, {Viti}, {Armus}, {Barr}, {Barcos-Mu{\~n}oz}, {Evans}, {Inami}, {Linden}, {Privon}, {Song}, {Rich}, {Aalto}, {Appleton}, {B{\"o}ker}, {Charmandaris}, {Diaz-Santos}, {Hayward}, {Lai}, {Medling}, {Ricci}, \& {U}}]{buiten24}
{Buiten}, V.~A., {van der Werf}, P.~P., {Viti}, S., {et~al.} 2024, \apj, 966, 166, \dodoi{10.3847/1538-4357/ad344b}

\bibitem[{{Calzetti} {et~al.}(2000){Calzetti}, {Armus}, {Bohlin}, {Kinney}, {Koornneef}, \& {Storchi-Bergmann}}]{calzetti00}
{Calzetti}, D., {Armus}, L., {Bohlin}, R.~C., {et~al.} 2000, \apj, 533, 682, \dodoi{10.1086/308692}

\bibitem[{{Cappellari}(2017)}]{cappellari17}
{Cappellari}, M. 2017, \mnras, 466, 798, \dodoi{10.1093/mnras/stw3020}

\bibitem[{{Charmandaris} {et~al.}(2004){Charmandaris}, {Le Floc'h}, \& {Mirabel}}]{charmandaris04}
{Charmandaris}, V., {Le Floc'h}, E., \& {Mirabel}, I.~F. 2004, \apjl, 600, L15, \dodoi{10.1086/381687}

\bibitem[{{Chevalier} \& {Clegg}(1985)}]{cc85}
{Chevalier}, R.~A., \& {Clegg}, A.~W. 1985, \nat, 317, 44, \dodoi{10.1038/317044a0}

\bibitem[{{Chu} \& {Kennicutt}(1994)}]{chukennicutt94}
{Chu}, Y.-H., \& {Kennicutt}, Jr., R.~C. 1994, \apj, 425, 720, \dodoi{10.1086/174017}

\bibitem[{{Dopita} \& {Sutherland}(1996)}]{dopitasutherland96}
{Dopita}, M.~A., \& {Sutherland}, R.~S. 1996, \apjs, 102, 161, \dodoi{10.1086/192255}

\bibitem[{{Dopita} \& {Sutherland}(2003)}]{dopita03}
---. 2003, {Astrophysics of the diffuse universe}, \dodoi{10.1007/978-3-662-05866-4}

\bibitem[{{Doyon} {et~al.}(1995){Doyon}, {Nadeau}, {Joseph}, {Goldader}, {Sanders}, \& {Rowlands}}]{Doyon95}
{Doyon}, R., {Nadeau}, D., {Joseph}, R.~D., {et~al.} 1995, \apj, 450, 111, \dodoi{10.1086/176123}

\bibitem[{{Dyson}(1979)}]{Dyson79}
{Dyson}, J.~E. 1979, \aap, 73, 132

\bibitem[{{Elmegreen} \& {Elmegreen}(1978)}]{elmegreen78}
{Elmegreen}, B.~G., \& {Elmegreen}, D.~M. 1978, \apj, 220, 1051, \dodoi{10.1086/155991}

\bibitem[{{Evans} {et~al.}(2022){Evans}, {Frayer}, {Charmandaris}, {Armus}, {Inami}, {Surace}, {Linden}, {Soifer}, {Diaz-Santos}, {Larson}, {Rich}, {Song}, {Barcos-Munoz}, {Mazzarella}, {Privon}, {U}, {Medling}, {B{\"o}ker}, {Aalto}, {Iwasawa}, {Howell}, {van der Werf}, {Appleton}, {Bohn}, {Brown}, {Hayward}, {Hoshioka}, {Kemper}, {Lai}, {Law}, {Malkan}, {Marshall}, {Murphy}, {Sanders}, \& {Stierwalt}}]{Evans22}
{Evans}, A.~S., {Frayer}, D.~T., {Charmandaris}, V., {et~al.} 2022, \apjl, 940, L8, \dodoi{10.3847/2041-8213/ac9971}

\bibitem[{{Franeck} {et~al.}(2022){Franeck}, {W{\"u}nsch}, {Mart{\'\i}nez-Gonz{\'a}lez}, {Orlitov{\'a}}, {Boorman}, {Svoboda}, {Sz{\'e}csi}, \& {Douna}}]{Franeck22}
{Franeck}, A., {W{\"u}nsch}, R., {Mart{\'\i}nez-Gonz{\'a}lez}, S., {et~al.} 2022, \apj, 927, 212, \dodoi{10.3847/1538-4357/ac4fc2}

\bibitem[{{Frayer} {et~al.}(1999){Frayer}, {Ivison}, {Smail}, {Yun}, \& {Armus}}]{Frayer99}
{Frayer}, D.~T., {Ivison}, R.~J., {Smail}, I., {Yun}, M.~S., \& {Armus}, L. 1999, \aj, 118, 139, \dodoi{10.1086/300925}

\bibitem[{{Goldader} {et~al.}(2002){Goldader}, {Meurer}, {Heckman}, {Seibert}, {Sanders}, {Calzetti}, \& {Steidel}}]{Goldader02}
{Goldader}, J.~D., {Meurer}, G., {Heckman}, T.~M., {et~al.} 2002, \apj, 568, 651, \dodoi{10.1086/339165}

\bibitem[{{Grimes} {et~al.}(2006){Grimes}, {Heckman}, {Hoopes}, {Strickland}, {Aloisi}, {Meurer}, \& {Ptak}}]{Grimes06}
{Grimes}, J.~P., {Heckman}, T., {Hoopes}, C., {et~al.} 2006, \apj, 648, 310, \dodoi{10.1086/505680}

\bibitem[{{Heckman} {et~al.}(1990){Heckman}, {Armus}, \& {Miley}}]{heckman_armus_miley90}
{Heckman}, T.~M., {Armus}, L., \& {Miley}, G.~K. 1990, \apjs, 74, 833, \dodoi{10.1086/191522}

\bibitem[{{Heckman} {et~al.}(2005){Heckman}, {Hoopes}, {Seibert}, {Martin}, {Salim}, {Rich}, {Kauffmann}, {Charlot}, {Barlow}, {Bianchi}, {Byun}, {Donas}, {Forster}, {Friedman}, {Jelinsky}, {Lee}, {Madore}, {Malina}, {Milliard}, {Morrissey}, {Neff}, {Schiminovich}, {Siegmund}, {Small}, {Szalay}, {Welsh}, \& {Wyder}}]{Heckman05}
{Heckman}, T.~M., {Hoopes}, C.~G., {Seibert}, M., {et~al.} 2005, \apjl, 619, L35, \dodoi{10.1086/425979}

\bibitem[{{Heiles}(1990)}]{heiles90}
{Heiles}, C. 1990, \apj, 354, 483, \dodoi{10.1086/168709}

\bibitem[{{Hopkins} {et~al.}(2013){Hopkins}, {Cox}, {Hernquist}, {Narayanan}, {Hayward}, \& {Murray}}]{hopkins13}
{Hopkins}, P.~F., {Cox}, T.~J., {Hernquist}, L., {et~al.} 2013, \mnras, 430, 1901, \dodoi{10.1093/mnras/stt017}

\bibitem[{{Howell} {et~al.}(2010){Howell}, {Armus}, {Mazzarella}, {Evans}, {Surace}, {Sanders}, {Petric}, {Appleton}, {Bothun}, {Bridge}, {Chan}, {Charmandaris}, {Frayer}, {Haan}, {Inami}, {Kim}, {Lord}, {Madore}, {Melbourne}, {Schulz}, {U}, {Vavilkin}, {Veilleux}, \& {Xu}}]{Howell10}
{Howell}, J.~H., {Armus}, L., {Mazzarella}, J.~M., {et~al.} 2010, \apj, 715, 572, \dodoi{10.1088/0004-637X/715/1/572}

\bibitem[{{Humire} {et~al.}(2025){Humire}, {Dey}, {Ronconi}, {Sasse}, {Cid Fernandes}, {Mart{\'\i}n}, {Donevski}, {Ma{\l}ek}, {Fern{\'a}ndez-Ontiveros}, {Song}, {Hamed}, {Mangum}, {Henkel}, {Rivilla}, {Colzi}, {Harada}, {Demarco}, {Goyal}, {Meier}, {Panda}, {Krabbe}, {Yan}, {Lopes}, {Sakamoto}, {Muller}, {Tanaka}, {Yoshimura}, {Nakanishi}, {Kanaan}, {Ribeiro}, {Schoenell}, \& {Mendes de Oliveira}}]{humire25}
{Humire}, P.~K., {Dey}, S., {Ronconi}, T., {et~al.} 2025, arXiv e-prints, arXiv:2501.15082, \dodoi{10.48550/arXiv.2501.15082}

\bibitem[{{Iono} {et~al.}(2004){Iono}, {Ho}, {Yun}, {Matsushita}, {Peck}, \& {Sakamoto}}]{iono04}
{Iono}, D., {Ho}, P. T.~P., {Yun}, M.~S., {et~al.} 2004, \apjl, 616, L63, \dodoi{10.1086/420784}

\bibitem[{Kass \& Raftery(1995)}]{kass95}
Kass, R.~E., \& Raftery, A.~E. 1995, Journal of the American Statistical Association, 90, 773, \dodoi{10.1080/01621459.1995.10476572}

\bibitem[{{Kauffmann} {et~al.}(2003){Kauffmann}, {Heckman}, {Tremonti}, {Brinchmann}, {Charlot}, {White}, {Ridgway}, {Brinkmann}, {Fukugita}, {Hall}, {Ivezi{\'c}}, {Richards}, \& {Schneider}}]{Kauffmann03}
{Kauffmann}, G., {Heckman}, T.~M., {Tremonti}, C., {et~al.} 2003, \mnras, 346, 1055, \dodoi{10.1111/j.1365-2966.2003.07154.x}

\bibitem[{{Kewley} {et~al.}(2001){Kewley}, {Dopita}, {Sutherland}, {Heisler}, \& {Trevena}}]{Kewley01}
{Kewley}, L.~J., {Dopita}, M.~A., {Sutherland}, R.~S., {Heisler}, C.~A., \& {Trevena}, J. 2001, \apj, 556, 121, \dodoi{10.1086/321545}

\bibitem[{{Kewley} {et~al.}(2006){Kewley}, {Groves}, {Kauffmann}, \& {Heckman}}]{Kewley06}
{Kewley}, L.~J., {Groves}, B., {Kauffmann}, G., \& {Heckman}, T. 2006, \mnras, 372, 961, \dodoi{10.1111/j.1365-2966.2006.10859.x}

\bibitem[{{Kinoshita} {et~al.}(2021{\natexlab{a}}){Kinoshita}, {Nakamura}, \& {Wu}}]{kinoshita21}
{Kinoshita}, S.~W., {Nakamura}, F., \& {Wu}, B. 2021{\natexlab{a}}, \apj, 921, 150, \dodoi{10.3847/1538-4357/ac1d4b}

\bibitem[{{Kinoshita} {et~al.}(2021{\natexlab{b}}){Kinoshita}, {Nakamura}, \& {Wu}}]{shinichi21}
---. 2021{\natexlab{b}}, arXiv e-prints, arXiv:2108.05554, \dodoi{10.48550/arXiv.2108.05554}

\bibitem[{{Knop} {et~al.}(1994){Knop}, {Soifer}, {Graham}, {Matthews}, {Sanders}, \& {Scoville}}]{Knop94}
{Knop}, R.~A., {Soifer}, B.~T., {Graham}, J.~R., {et~al.} 1994, \aj, 107, 920, \dodoi{10.1086/116906}

\bibitem[{{Kreckel} {et~al.}(2014){Kreckel}, {Armus}, {Groves}, {Lyubenova}, {D{\'\i}az-Santos}, {Schinnerer}, {Appleton}, {Croxall}, {Dale}, {Hunt}, {Beir{\~a}o}, {Bolatto}, {Calzetti}, {Donovan Meyer}, {Draine}, {Hinz}, {Kennicutt}, {Meidt}, {Murphy}, {Smith}, {Tabatabaei}, \& {Walter}}]{kreckel14}
{Kreckel}, K., {Armus}, L., {Groves}, B., {et~al.} 2014, \apj, 790, 26, \dodoi{10.1088/0004-637X/790/1/26}

\bibitem[{{Larkin} {et~al.}(1998){Larkin}, {Armus}, {Knop}, {Soifer}, \& {Matthews}}]{Larkin88}
{Larkin}, J.~E., {Armus}, L., {Knop}, R.~A., {Soifer}, B.~T., \& {Matthews}, K. 1998, \apjs, 114, 59, \dodoi{10.1086/313063}

\bibitem[{{Larson} {et~al.}(2020){Larson}, {D{\'\i}az-Santos}, {Armus}, {Privon}, {Linden}, {Evans}, {Howell}, {Charmandaris}, {U}, {Sanders}, {Stierwalt}, {Barcos-Mu{\~n}oz}, {Rich}, {Medling}, {Cook}, {Oklop{\^{c}}i{\'c}}, {Murphy}, \& {Bonfini}}]{larson20}
{Larson}, K.~L., {D{\'\i}az-Santos}, T., {Armus}, L., {et~al.} 2020, \apj, 888, 92, \dodoi{10.3847/1538-4357/ab5dc3}

\bibitem[{{Le Floc'h} {et~al.}(2002){Le Floc'h}, {Charmandaris}, {Laurent}, {Mirabel}, {Gallais}, {Sauvage}, {Vigroux}, \& {Cesarsky}}]{LeFloch02}
{Le Floc'h}, E., {Charmandaris}, V., {Laurent}, O., {et~al.} 2002, \aap, 391, 417, \dodoi{10.1051/0004-6361:20020784}

\bibitem[{{Leitherer} {et~al.}(1999){Leitherer}, {Schaerer}, {Goldader}, {Delgado}, {Robert}, {Kune}, {de Mello}, {Devost}, \& {Heckman}}]{leitherer99}
{Leitherer}, C., {Schaerer}, D., {Goldader}, J.~D., {et~al.} 1999, \apjs, 123, 3, \dodoi{10.1086/313233}

\bibitem[{{Linden} {et~al.}(2021){Linden}, {Evans}, {Larson}, {Privon}, {Armus}, {Rich}, {D{\'\i}az-Santos}, {Murphy}, {Song}, {Barcos-Mu{\~n}oz}, {Howell}, {Charmandaris}, {Inami}, {U}, {Surace}, {Mazzarella}, \& {Calzetti}}]{linden21}
{Linden}, S.~T., {Evans}, A.~S., {Larson}, K., {et~al.} 2021, \apj, 923, 278, \dodoi{10.3847/1538-4357/ac2892}

\bibitem[{{Linden} {et~al.}(2023){Linden}, {Evans}, {Armus}, {Rich}, {Larson}, {Lai}, {Privon}, {U}, {Inami}, {Bohn}, {Song}, {Barcos-Mu{\~n}oz}, {Charmandaris}, {Medling}, {Stierwalt}, {Diaz-Santos}, {B{\"o}ker}, {van der Werf}, {Aalto}, {Appleton}, {Brown}, {Hayward}, {Howell}, {Iwasawa}, {Kemper}, {Frayer}, {Law}, {Malkan}, {Marshall}, {Mazzarella}, {Murphy}, {Sanders}, \& {Surace}}]{linden23}
{Linden}, S.~T., {Evans}, A.~S., {Armus}, L., {et~al.} 2023, \apjl, 944, L55, \dodoi{10.3847/2041-8213/acb335}

\bibitem[{{Luridiana} {et~al.}(2013){Luridiana}, {Morisset}, \& {Shaw}}]{luridiana13}
{Luridiana}, V., {Morisset}, C., \& {Shaw}, R.~A. 2013, {PyNeb: Analysis of emission lines}, Astrophysics Source Code Library, record ascl:1304.021

\bibitem[{{Mac Low} \& {Ferrara}(1999)}]{maclow99}
{Mac Low}, M.-M., \& {Ferrara}, A. 1999, \apj, 513, 142, \dodoi{10.1086/306832}

\bibitem[{{Medling} {et~al.}(2015){Medling}, {U}, {Rich}, {Kewley}, {Armus}, {Dopita}, {Max}, {Sanders}, \& {Sutherland}}]{medling15}
{Medling}, A.~M., {U}, V., {Rich}, J.~A., {et~al.} 2015, \mnras, 448, 2301, \dodoi{10.1093/mnras/stv081}

\bibitem[{{Melnick}(1979)}]{melnick79}
{Melnick}, J. 1979, \apj, 228, 112, \dodoi{10.1086/156827}

\bibitem[{{Melnick} {et~al.}(2021){Melnick}, {Tenorio-Tagle}, \& {Telles}}]{melnick21}
{Melnick}, J., {Tenorio-Tagle}, G., \& {Telles}, E. 2021, \aap, 649, A175, \dodoi{10.1051/0004-6361/201937268}

\bibitem[{{Morisset} \& {Delgado-Inglada}(2014)}]{morisset14}
{Morisset}, C., \& {Delgado-Inglada}, G. 2014, in Revista Mexicana de Astronomia y Astrofisica Conference Series, Vol.~44, Revista Mexicana de Astronomia y Astrofisica Conference Series, 136--136

\bibitem[{{Morrissey} {et~al.}(2018){Morrissey}, {Matuszewski}, {Martin}, {Neill}, {Epps}, {Fucik}, {Weber}, {Darvish}, {Adkins}, {Allen}, {Bartos}, {Belicki}, {Cabak}, {Callahan}, {Cowley}, {Crabill}, {Deich}, {Delecroix}, {Doppman}, {Hilyard}, {James}, {Kaye}, {Kokorowski}, {Kwok}, {Lanclos}, {Milner}, {Moore}, {O'Sullivan}, {Parihar}, {Park}, {Phillips}, {Rizzi}, {Rockosi}, {Rodriguez}, {Salaun}, {Seaman}, {Sheikh}, {Weiss}, \& {Zarzaca}}]{morrisey18}
{Morrissey}, P., {Matuszewski}, M., {Martin}, D.~C., {et~al.} 2018, \apj, 864, 93, \dodoi{10.3847/1538-4357/aad597}

\bibitem[{{Munoz-Tunon} {et~al.}(1996){Munoz-Tunon}, {Tenorio-Tagle}, {Castaneda}, \& {Terlevich}}]{MunozTunon96}
{Munoz-Tunon}, C., {Tenorio-Tagle}, G., {Castaneda}, H.~O., \& {Terlevich}, R. 1996, \aj, 112, 1636, \dodoi{10.1086/118129}

\bibitem[{{Murray} {et~al.}(2011){Murray}, {M{\'e}nard}, \& {Thompson}}]{murray11}
{Murray}, N., {M{\'e}nard}, B., \& {Thompson}, T.~A. 2011, \apj, 735, 66, \dodoi{10.1088/0004-637X/735/1/66}

\bibitem[{{Naab} \& {Ostriker}(2017)}]{naabandostriker17}
{Naab}, T., \& {Ostriker}, J.~P. 2017, \araa, 55, 59, \dodoi{10.1146/annurev-astro-081913-040019}

\bibitem[{{Osterbrock} \& {Ferland}(2006)}]{osterbrockferland06}
{Osterbrock}, D.~E., \& {Ferland}, G.~J. 2006, {Astrophysics of gaseous nebulae and active galactic nuclei}

\bibitem[{{O'Sullivan} \& {Chen}(2020)}]{osullivan20}
{O'Sullivan}, D., \& {Chen}, Y. 2020, arXiv e-prints, arXiv:2011.05444, \dodoi{10.48550/arXiv.2011.05444}

\bibitem[{{Ot{\'\i}-Floranes} \& {Mas-Hesse}(2010)}]{OtiFloranes10}
{Ot{\'\i}-Floranes}, H., \& {Mas-Hesse}, J.~M. 2010, \aap, 511, A61, \dodoi{10.1051/0004-6361/200913384}

\bibitem[{{Reichardt Chu} {et~al.}(2022){Reichardt Chu}, {Fisher}, {Nielsen}, {Chisholm}, {Girard}, {Kacprzak}, {Bolatto}, {Herrera-Camus}, {Sandstrom}, {Li}, {Rickards Vaught}, \& {McPherson}}]{rehichardtchu22}
{Reichardt Chu}, B., {Fisher}, D.~B., {Nielsen}, N.~M., {et~al.} 2022, \mnras, 511, 5782, \dodoi{10.1093/mnras/stac420}

\bibitem[{{Rich} {et~al.}(2023){Rich}, {Aalto}, {Evans}, {Charmandaris}, {Privon}, {Lai}, {Inami}, {Linden}, {Armus}, {Diaz-Santos}, {Appleton}, {Barcos-Mu{\~n}oz}, {B{\"o}ker}, {Larson}, {Law}, {Malkan}, {Medling}, {Song}, {U}, {van der Werf}, {Bohn}, {Brown}, {Finnerty}, {Hayward}, {Howell}, {Iwasawa}, {Kemper}, {Marshall}, {Mazzarella}, {McKinney}, {Muller-Sanchez}, {Murphy}, {Sanders}, {Soifer}, {Stierwalt}, \& {Surace}}]{Rich23}
{Rich}, J., {Aalto}, S., {Evans}, A.~S., {et~al.} 2023, \apjl, 944, L50, \dodoi{10.3847/2041-8213/acb2b8}

\bibitem[{{Rich} {et~al.}(2010){Rich}, {Dopita}, {Kewley}, \& {Rupke}}]{rich10}
{Rich}, J.~A., {Dopita}, M.~A., {Kewley}, L.~J., \& {Rupke}, D.~S.~N. 2010, \apj, 721, 505, \dodoi{10.1088/0004-637X/721/1/505}

\bibitem[{{Rich} {et~al.}(2011){Rich}, {Kewley}, \& {Dopita}}]{rich11}
{Rich}, J.~A., {Kewley}, L.~J., \& {Dopita}, M.~A. 2011, \apj, 734, 87, \dodoi{10.1088/0004-637X/734/2/87}

\bibitem[{{Rich} {et~al.}(2015){Rich}, {Kewley}, \& {Dopita}}]{rich15}
---. 2015, \apjs, 221, 28, \dodoi{10.1088/0067-0049/221/2/28}

\bibitem[{{Riffel} {et~al.}(2013){Riffel}, {Rodr{\'\i}guez-Ardila}, {Aleman}, {Brotherton}, {Pastoriza}, {Bonatto}, \& {Dors}}]{Riffel13}
{Riffel}, R., {Rodr{\'\i}guez-Ardila}, A., {Aleman}, I., {et~al.} 2013, \mnras, 430, 2002, \dodoi{10.1093/mnras/stt026}

\bibitem[{{Riffel} {et~al.}(2021){Riffel}, {Bianchin}, {Riffel}, {Storchi-Bergmann}, {Sch{\"o}nell}, {Dahmer-Hahn}, {Dametto}, \& {Diniz}}]{Riffel21}
{Riffel}, R.~A., {Bianchin}, M., {Riffel}, R., {et~al.} 2021, \mnras, 503, 5161, \dodoi{10.1093/mnras/stab788}

\bibitem[{{Saito} {et~al.}(2015){Saito}, {Iono}, {Yun}, {Ueda}, {Nakanishi}, {Sugai}, {Espada}, {Imanishi}, {Motohara}, {Hagiwara}, {Tateuchi}, {Lee}, \& {Kawabe}}]{saito15}
{Saito}, T., {Iono}, D., {Yun}, M.~S., {et~al.} 2015, \apj, 803, 60, \dodoi{10.1088/0004-637X/803/2/60}

\bibitem[{{Saito} {et~al.}(2017){Saito}, {Iono}, {Espada}, {Nakanishi}, {Ueda}, {Sugai}, {Takano}, {Yun}, {Imanishi}, {Ohashi}, {Lee}, {Hagiwara}, {Motohara}, \& {Kawabe}}]{saito17}
{Saito}, T., {Iono}, D., {Espada}, D., {et~al.} 2017, \apj, 834, 6, \dodoi{10.3847/1538-4357/834/1/6}

\bibitem[{{Saito} {et~al.}(2018){Saito}, {Iono}, {Espada}, {Nakanishi}, {Ueda}, {Sugai}, {Yun}, {Takano}, {Imanishi}, {Michiyama}, {Ohashi}, {Lee}, {Hagiwara}, {Motohara}, {Yamashita}, {Ando}, \& {Kawabe}}]{saito18}
---. 2018, \apj, 863, 129, \dodoi{10.3847/1538-4357/aad23b}

\bibitem[{{Sanders} {et~al.}(2003){Sanders}, {Mazzarella}, {Kim}, {Surace}, \& {Soifer}}]{sanders03}
{Sanders}, D.~B., {Mazzarella}, J.~M., {Kim}, D.~C., {Surace}, J.~A., \& {Soifer}, B.~T. 2003, \aj, 126, 1607, \dodoi{10.1086/376841}

\bibitem[{{Schaye} {et~al.}(2015){Schaye}, {Crain}, {Bower}, {Furlong}, {Schaller}, {Theuns}, {Dalla Vecchia}, {Frenk}, {McCarthy}, {Helly}, {Jenkins}, {Rosas-Guevara}, {White}, {Baes}, {Booth}, {Camps}, {Navarro}, {Qu}, {Rahmati}, {Sawala}, {Thomas}, \& {Trayford}}]{schaye15}
{Schaye}, J., {Crain}, R.~A., {Bower}, R.~G., {et~al.} 2015, \mnras, 446, 521, \dodoi{10.1093/mnras/stu2058}

\bibitem[{{Scoville} {et~al.}(2000){Scoville}, {Evans}, {Thompson}, {Rieke}, {Hines}, {Low}, {Dinshaw}, {Surace}, \& {Armus}}]{Scoville00}
{Scoville}, N.~Z., {Evans}, A.~S., {Thompson}, R., {et~al.} 2000, \aj, 119, 991, \dodoi{10.1086/301248}

\bibitem[{{Sexton} {et~al.}(2021){Sexton}, {Matzko}, {Darden}, {Canalizo}, \& {Gorjian}}]{sexton21}
{Sexton}, R.~O., {Matzko}, W., {Darden}, N., {Canalizo}, G., \& {Gorjian}, V. 2021, \mnras, 500, 2871, \dodoi{10.1093/mnras/staa3278}

\bibitem[{{Sharp} \& {Bland-Hawthorn}(2010)}]{sharp10}
{Sharp}, R.~G., \& {Bland-Hawthorn}, J. 2010, \apj, 711, 818, \dodoi{10.1088/0004-637X/711/2/818}

\bibitem[{{Shopbell} \& {Bland-Hawthorn}(1998)}]{shopbell98}
{Shopbell}, P.~L., \& {Bland-Hawthorn}, J. 1998, \apj, 493, 129, \dodoi{10.1086/305108}

\bibitem[{{Silich} {et~al.}(2004){Silich}, {Tenorio-Tagle}, \& {Rodr{\'\i}guez-Gonz{\'a}lez}}]{silich04}
{Silich}, S., {Tenorio-Tagle}, G., \& {Rodr{\'\i}guez-Gonz{\'a}lez}, A. 2004, \apj, 610, 226, \dodoi{10.1086/421702}

\bibitem[{{Slavin} {et~al.}(1993){Slavin}, {Shull}, \& {Begelman}}]{slavin93}
{Slavin}, J.~D., {Shull}, J.~M., \& {Begelman}, M.~C. 1993, \apj, 407, 83, \dodoi{10.1086/172494}

\bibitem[{{Soifer} {et~al.}(1987){Soifer}, {Neugebauer}, \& {Houck}}]{Soifer87}
{Soifer}, B.~T., {Neugebauer}, G., \& {Houck}, J.~R. 1987, \araa, 25, 187, \dodoi{10.1146/annurev.aa.25.090187.001155}

\bibitem[{{Soifer} {et~al.}(2001){Soifer}, {Neugebauer}, {Matthews}, {Egami}, {Weinberger}, {Ressler}, {Scoville}, {Stolovy}, {Condon}, \& {Becklin}}]{Soifer01}
{Soifer}, B.~T., {Neugebauer}, G., {Matthews}, K., {et~al.} 2001, \aj, 122, 1213, \dodoi{10.1086/322119}

\bibitem[{{Song} {et~al.}(2022){Song}, {Linden}, {Evans}, {Barcos-Mu{\~n}oz}, {Murphy}, {Momjian}, {D{\'\i}az-Santos}, {Larson}, {Privon}, {Huang}, {Armus}, {Mazzarella}, {U}, {Inami}, {Charmandaris}, {Ricci}, {Emig}, {McKinney}, {Yoon}, {Kunneriath}, {Lai}, {Rodas-Quito}, {Saravia}, {Gao}, {Meynardie}, \& {Sanders}}]{song22}
{Song}, Y., {Linden}, S.~T., {Evans}, A.~S., {et~al.} 2022, \apj, 940, 52, \dodoi{10.3847/1538-4357/ac923b}

\bibitem[{{Stevens} \& {Hartwell}(2003)}]{stevens_hartwell03}
{Stevens}, I.~R., \& {Hartwell}, J.~M. 2003, \mnras, 339, 280, \dodoi{10.1046/j.1365-8711.2003.06184.x}

\bibitem[{{Sutherland} \& {Dopita}(2017)}]{sutherlanddopita17}
{Sutherland}, R.~S., \& {Dopita}, M.~A. 2017, \apjs, 229, 34, \dodoi{10.3847/1538-4365/aa6541}

\bibitem[{{Swinbank} {et~al.}(2019){Swinbank}, {Harrison}, {Tiley}, {Johnson}, {Smail}, {Stott}, {Best}, {Bower}, {Bureau}, {Bunker}, {Cirasuolo}, {Jarvis}, {Magdis}, {Sharples}, \& {Sobral}}]{swinbank19}
{Swinbank}, M., {Harrison}, C., {Tiley}, A., {et~al.} 2019, arXiv e-prints, arXiv:1906.05311, \dodoi{10.48550/arXiv.1906.05311}

\bibitem[{{Tenorio-Tagle} {et~al.}(1996){Tenorio-Tagle}, {Munoz-Tunon}, \& {Cid-Fernandes}}]{tenoriotagle96}
{Tenorio-Tagle}, G., {Munoz-Tunon}, C., \& {Cid-Fernandes}, R. 1996, \apj, 456, 264, \dodoi{10.1086/176646}

\bibitem[{{Tenorio-Tagle} {et~al.}(1993){Tenorio-Tagle}, {Munoz-Tunon}, \& {Cox}}]{tenoriotagle93}
{Tenorio-Tagle}, G., {Munoz-Tunon}, C., \& {Cox}, D.~P. 1993, \apj, 418, 767, \dodoi{10.1086/173434}

\bibitem[{{Terlevich} \& {Melnick}(1981)}]{terlevich81}
{Terlevich}, R., \& {Melnick}, J. 1981, \mnras, 195, 839, \dodoi{10.1093/mnras/195.4.839}

\bibitem[{{Valdes} {et~al.}(2004){Valdes}, {Gupta}, {Rose}, {Singh}, \& {Bell}}]{valdes04}
{Valdes}, F., {Gupta}, R., {Rose}, J.~A., {Singh}, H.~P., \& {Bell}, D.~J. 2004, \apjs, 152, 251, \dodoi{10.1086/386343}

\bibitem[{{van Driel} {et~al.}(2001){van Driel}, {Gao}, \& {Monnier-Ragaigne}}]{vanDriel01}
{van Driel}, W., {Gao}, Y., \& {Monnier-Ragaigne}, D. 2001, \aap, 368, 64, \dodoi{10.1051/0004-6361:20000509}

\bibitem[{{Veilleux} {et~al.}(2005){Veilleux}, {Cecil}, \& {Bland-Hawthorn}}]{veilleux05}
{Veilleux}, S., {Cecil}, G., \& {Bland-Hawthorn}, J. 2005, \araa, 43, 769, \dodoi{10.1146/annurev.astro.43.072103.150610}

\bibitem[{{Veilleux} \& {Osterbrock}(1987)}]{Veilleux87}
{Veilleux}, S., \& {Osterbrock}, D.~E. 1987, \apjs, 63, 295, \dodoi{10.1086/191166}

\bibitem[{{Vogelsberger} {et~al.}(2020){Vogelsberger}, {Marinacci}, {Torrey}, \& {Puchwein}}]{vogelsberger20}
{Vogelsberger}, M., {Marinacci}, F., {Torrey}, P., \& {Puchwein}, E. 2020, Nature Reviews Physics, 2, 42, \dodoi{10.1038/s42254-019-0127-2}

\bibitem[{{Westmoquette} {et~al.}(2012){Westmoquette}, {Clements}, {Bendo}, \& {Khan}}]{westmoquette12}
{Westmoquette}, M.~S., {Clements}, D.~L., {Bendo}, G.~J., \& {Khan}, S.~A. 2012, \mnras, 424, 416, \dodoi{10.1111/j.1365-2966.2012.21214.x}

\bibitem[{{Westmoquette} {et~al.}(2011){Westmoquette}, {Smith}, \& {Gallagher}}]{westmoquette11}
{Westmoquette}, M.~S., {Smith}, L.~J., \& {Gallagher}, III, J.~S. 2011, \mnras, 414, 3719, \dodoi{10.1111/j.1365-2966.2011.18675.x}

\bibitem[{{W{\"u}nsch} {et~al.}(2007){W{\"u}nsch}, {Silich}, {Palou{\v{s}}}, \& {Tenorio-Tagle}}]{wunsch07}
{W{\"u}nsch}, R., {Silich}, S., {Palou{\v{s}}}, J., \& {Tenorio-Tagle}, G. 2007, \aap, 471, 579, \dodoi{10.1051/0004-6361:20077282}

\bibitem[{{W{\"u}nsch} {et~al.}(2011){W{\"u}nsch}, {Silich}, {Palou{\v{s}}}, {Tenorio-Tagle}, \& {Mu{\~n}oz-Tu{\~n}{\'o}n}}]{wunsch11}
{W{\"u}nsch}, R., {Silich}, S., {Palou{\v{s}}}, J., {Tenorio-Tagle}, G., \& {Mu{\~n}oz-Tu{\~n}{\'o}n}, C. 2011, \apj, 740, 75, \dodoi{10.1088/0004-637X/740/2/75}

\bibitem[{{Yun} {et~al.}(1994){Yun}, {Scoville}, \& {Knop}}]{Yun94}
{Yun}, M.~S., {Scoville}, N.~Z., \& {Knop}, R.~A. 1994, \apjl, 430, L109, \dodoi{10.1086/187450}

\end{thebibliography}
\bibliographystyle{aasjournal}

\appendix
\setcounter{figure}{0} % Reset figure counter for appendix
\renewcommand{\thefigure}{A\arabic{figure}} % Set appendix figure numbering to A1, A2, ...
\setcounter{table}{0}  % Reset table counter for appendix
\renewcommand{\thetable}{A\arabic{table}}  % Set appendix table numbering to A1, A2, ...

\section{VV 114 Atlas}
\label{appendixA}

Throughout the paper, various regions of VV 114 are referenced. For convenience, this appendix contains a \emph{JWST} MIRI F770W image of VV 114 with labeled boxes indicating major structures of the galaxy. The black contours are the same as in Figure \ref{Figure_RGB}. 

\begin{figure}[h!]
    \centering
    \includegraphics[width=0.47\textwidth]{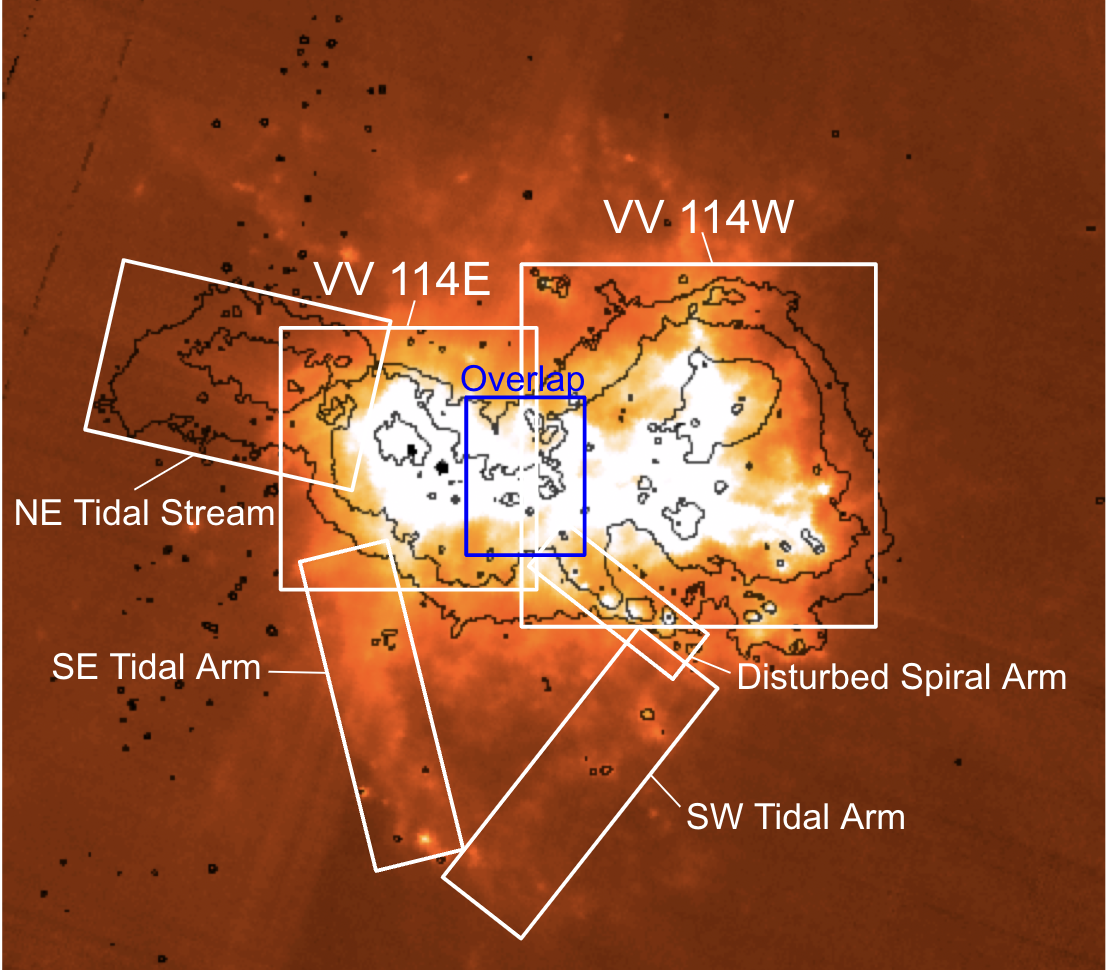}
    \caption{\emph{JWST} MIRI F770W image of VV 114 with markers indicating prominent regions visible in the optical and mid-IR wavebands. The black contours are the same as shown in Figure \ref{Figure_RGB}.}
    \label{Figure_Atlas}
\end{figure}

\section{Emission Line Ratio Maps}
\label{appendixB}

Emission line ratios from the optical MUSE IFU data (\nii/H$\alpha$ and \sii/H$\alpha$) and from the near-IR \emph{HST} imaging data (\feii/\pab) are shown in this appendix section for reference. The optical images are masked to include only spaxels where \nii, \sii, and H$\alpha$ had SNR $>$ 10, and the near-IR image was limited to where the flux density was $ f_{\rm{\lambda}}\geq 3\sigma$~based on pixel values in background portions of the image far from the galaxy. The line images were masked to include only portions of the galaxy where SNR $>$ 10 for all five emission lines.

\begin{figure}[h!]
    \centering
    \includegraphics[width=\textwidth]{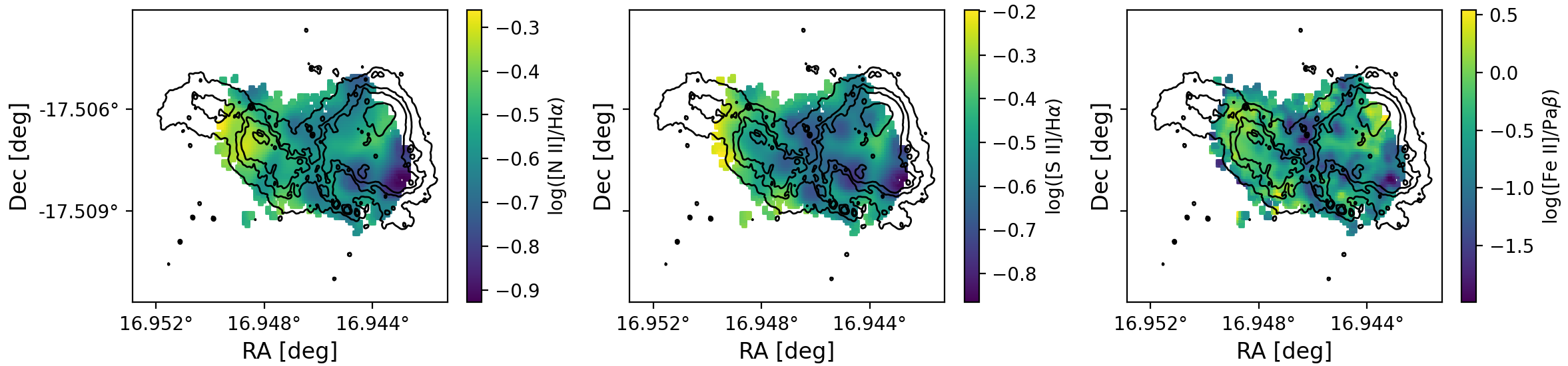}
    \caption{Left to right: maps of the logarithmic \nii/H$\alpha$, \sii/H$\alpha$, and \feii/\pab~emission line ratios. The line ratio images are individually scaled to the 3-$\sigma$ envelope around the pixel median value to highlight the similarity in spatial trends between the images. Contours are the same as Figure \ref{Figure_RGB}.}
    \label{Figure_NIRratiomaps}
\end{figure}

\section{Cluster Wind Model Fits to star clusters}
\label{appendixC}

Presented here are the cluster wind model fits to the radial electron density profiles of the gas surrounding 66 star clusters in VV 114, as well as the correlation that was found between the fitted values of $\dot{E}_{\rm{T}}$ and $R_{\rm{sc}}$. The best-fit parameters for each cluster can be found in Table \ref{Table1}. Panels have identical ranges in logarithmic radius (0.5 $<$~log($d_{\rm{cluster}} \rm{[pc]}$) $<$~4) and electron density (-0.5 $<$~log($N_{\rm{e}} \rm{[cm^{-3}]})$~$<$~3.5). For each cluster, the measurements that comprise the radial density profile come from all those spaxels closer to that cluster than to any other, which explains the variation in the number of points in each plot. Also included in this Appendix section is the correlation between the kinetic energy injection rate $\dot{E}_{\rm{T}}$ and injection zone radius $R_{\rm{sc}}$ among the 66 cluster wind fits (Figure \ref{Figure_WindParamCorrelation}.)

\begin{figure*}[ht!]
    \centering
    \includegraphics[width=\textwidth]{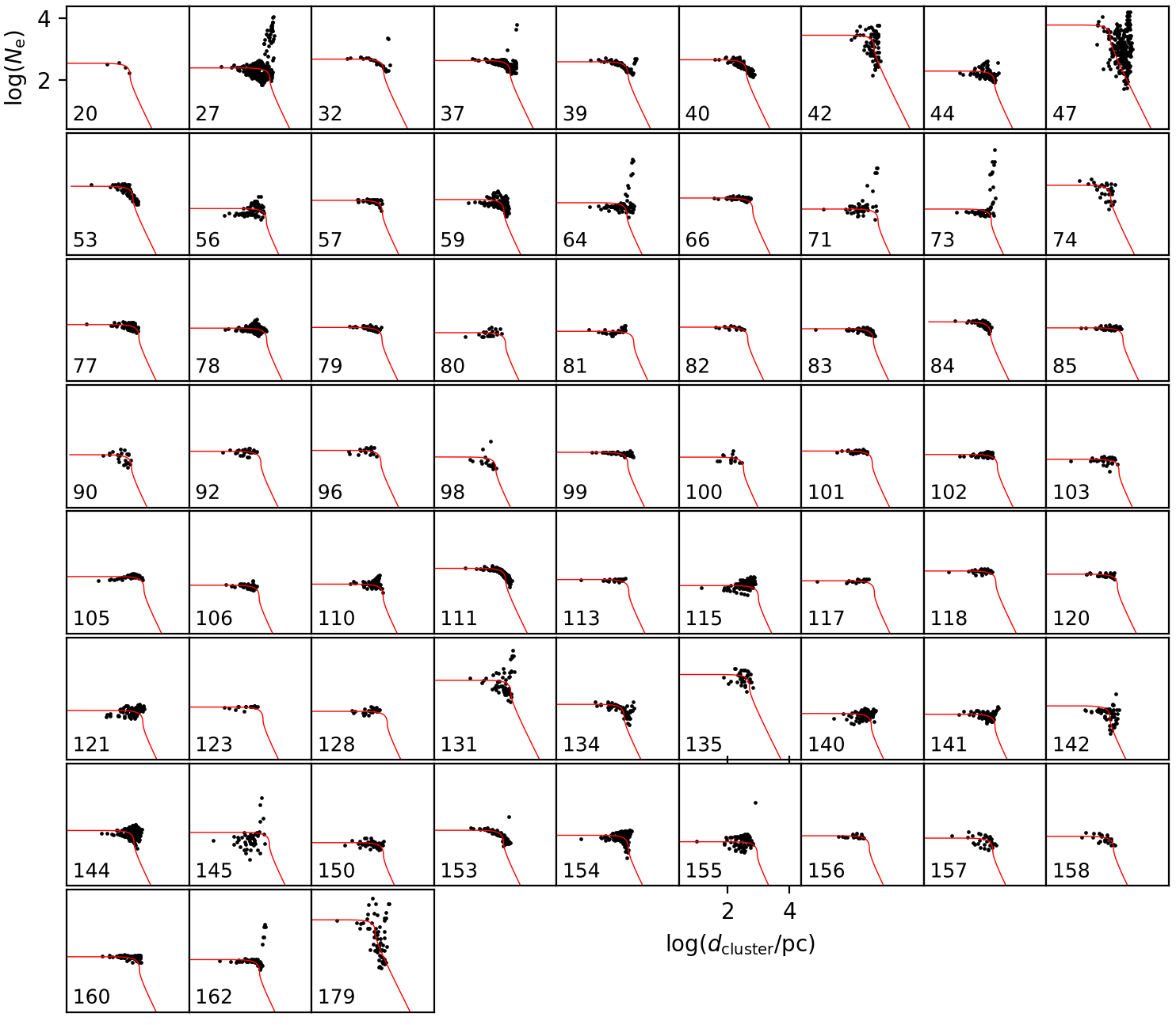}
    \caption{Radial electron density profiles of ionized gas surrounding the various star clusters in VV 114 (black points) fitted with the CC85 cluster winds model (red curves.) Cluster ID numbers come from \cite{linden21}.}
    \label{Figure_ClusterWindFits}
\end{figure*}

\begin{figure*}[ht!]
    \centering
    \includegraphics[width=0.5\textwidth]{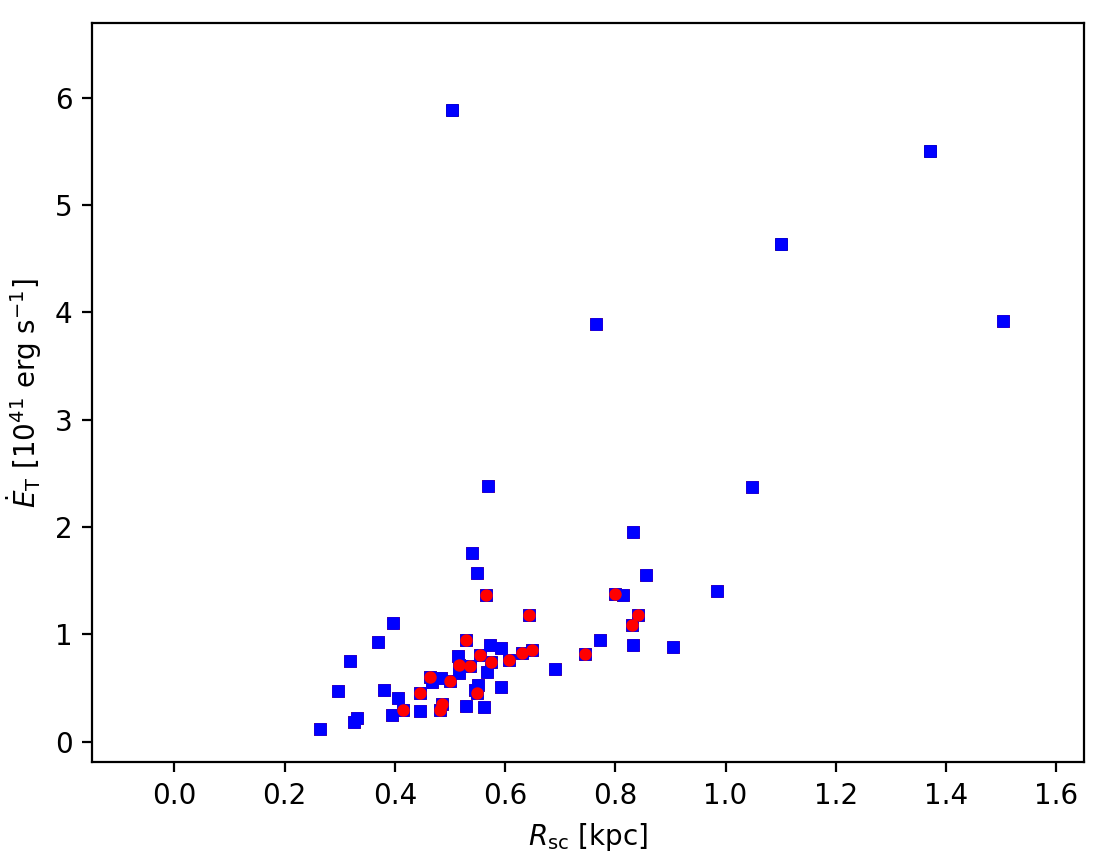}
    \caption{There is a moderate correlation between the energy injection rate $\dot{E}_{\rm{T}}$ and injection radius $R_{\rm{sc}}$, excluding the 5 clusters with $\dot{E}_{\rm{T}} > 6.7\times 10^{41}$~erg s$^{-1}$, with a Pearson correlation coefficient of $r_{\rm{Pearson}}$ = 0.64; $p$-value = 3$\times$10$^{-8}$ (blue points). The red points are best fitting wind model parameters for clusters showing flat density profiles, where we find a stronger correlation between $\dot{E}_{\rm{T}}$ and $R_{\rm{sc}}$: $r_{\rm{Pearson}}$ = 0.74; $p$-value = 10$^{-4}$. The parameters are correlated despite not being tied in the model fitting process, suggesting the cluster wind model is a reasonable description of the radial density profile data.}
    \label{Figure_WindParamCorrelation}
\end{figure*}

\begin{table*}[ht]
\centering
\caption{Derived properties of individual star clusters in VV 114. Inequality symbols indicate the fitted parameters are lower limits. (Continued from Table \ref{Table1}.)}
\label{Table A1}
\begin{tabular}{llllllll}
\hline
\hline
ID$^{a}$ & R.A. {[}deg{]} & Dec {[}deg{]} & log($M/M_{\odot}$)$^{a}$ & log($\alpha$/yr)$^{a}$ & $R_{\rm{sc}}$ {[}kpc{]} & $\dot{E}_{\rm{T}}$ {[}10$^{41}$~erg s$^{-1}${]} \\
\hline
56     & 16.946442      & -17.506558     & 8.19667                & 8.70816              & $\geq$0.83                    & $\geq$0.90                                           \\
57     & 16.946103      & -17.506868     & 5.50136                & 6.61453              & 0.52                    & 0.64                                           \\
59     & 16.946966      & -17.507904     & 4.25439                & 6.97089              & 0.48                    & 0.59                                           \\
64     & 16.946698      & -17.508163     & 5.02657                & 6.4809               & $\geq$1.34                    & $\geq$6.96                                           \\
66     & 16.945989      & -17.506781     & 4.82115                & 7.23817              & $\geq$0.64                    & $\geq$1.18                                           \\
71     & 16.946339      & -17.508488     & 5.50141                & 6.4809               & $\geq$1.37                    & $\geq$5.51                                           \\
73     & 16.945421      & -17.506156     & 5.09149                & 7.32726              & $\geq$0.83                    & $\geq$1.09                                           \\
74     & 16.944432      & -17.506156     & 5.19157                & 6.30272              & 0.32                    & 0.75                                           \\
77     & 16.946048      & -17.507362     & 6.48685                & 7.41634              & 0.57                    & 0.90                                           \\
78     & 16.946077      & -17.50782      & 6.81534                & 8.73044              & $\geq$0.86                    & $\geq$1.55                                           \\
79     & 16.945831      & -17.506947     & 5.31299                & 6.70362              & $\geq$0.50                    & $\geq$0.56                                           \\
80     & 16.945393      & -17.506409     & 4.95967                & 7.01544              & $\geq$0.55                    & $\geq$0.45                                           \\
81     & 16.945539      & -17.506419     & 5.38145                & 7.46088              & $\geq$0.84                    & $\geq$1.18                                           \\
82     & 16.94596       & -17.507268     & 5.45724                & 6.01317              & $\geq$0.45                    & $\geq$0.45                                           \\
83     & 16.946164      & -17.508365     & 5.03922                & 6.4809               & 0.57                    & 0.65                                           \\
84     & 16.945128      & -17.506544     & 5.90003                & 7.23817              & 0.38                    & 0.48                                           \\
85     & 16.945873      & -17.50728      & 5.84428                & 7.26043              & $\geq$0.80                    & $\geq$1.38                                           \\
90     & 16.944782      & -17.506406     & 4.74754                & 6.70362              & 0.33                    & 0.22                                           \\
92     & 16.944784      & -17.506763     & 4.86946                & 6.70362              & $\geq$0.55                    & $\geq$0.81                                           \\
96     & 16.944851      & -17.506476     & 4.53686                & 6.61453              & $\geq$0.46                    & $\geq$0.61                                           \\
98     & 16.944592      & -17.506061     & 5.16999                & 8.06224              & 0.26                    & 0.12                                           \\
99     & 16.945253      & -17.507264     & 5.78702                & 6.4809               & $\geq$0.54                    & $\geq$0.71                                           \\
100    & 16.944724      & -17.506559     & 4.74227                & 6.97089              & $\geq$0.33                    & $\geq$0.18                                           \\
101    & 16.944855      & -17.507322     & 5.95446                & 6.4809               & $\geq$0.52                    & $\geq$0.72                                           \\
102    & 16.94532       & -17.506976     & 6.00108                & 6.30272              & $\geq$0.63                    & $\geq$0.83                                           \\
103    & 16.946092      & -17.508643     & 5.97105                & 6.30272              & $\geq$0.59                    & $\geq$0.51                                           \\
105    & 16.945096      & -17.507322     & 4.91831                & 7.01544              & $\geq$0.83                    & $\geq$1.96                                           \\
106    & 16.944492      & -17.506629     & 5.51087                & 8.44089              & $\geq$0.49                    & $\geq$0.35                                           \\
110    & 16.944214      & -17.507279     & 4.38643                & 6.61453              & 0.55                    & 0.48                                           \\
111    & 16.94478       & -17.507754     & 5.21742                & 6.01317              & 0.55                    & 1.57                                           \\
113    & 16.945057      & -17.507268     & 6.35229                & 7.14907              & $\geq$0.57                    & $\geq$0.75                                           \\
115    & 16.945353      & -17.508308     & 5.5505                 & 8.41862              & $\geq$0.98                    & $\geq$1.40
\\
\hline
\multicolumn{4}{l}{\small {\bf{Note.}} $^a$ values taken from \cite{linden21}.} \\
\end{tabular}
\end{table*}

\begin{table*}[ht]
\centering
\caption{Derived properties of individual star clusters in VV 114. Inequality symbols indicate the fitted parameters are lower limits. (Continued from Table \ref{Table A1}.)}
\label{Table A2}
\begin{tabular}{llllllll}
\hline
\hline
ID$^{a}$ & R.A. {[}deg{]} & Dec {[}deg{]} & log($M/M_{\odot}$)$^{a}$ & log($\alpha$/yr)$^{a}$ & $R_{\rm{sc}}$ {[}kpc{]} & $\dot{E}_{\rm{T}}$ {[}10$^{41}$~erg s$^{-1}${]} \\
\hline
117    & 16.944648      & -17.506919     & 5.77168                & 8.17362              & $\geq$0.61                    & $\geq$0.76                                           \\
118    & 16.944752      & -17.507531     & 4.45422                & 6.4809               & $\geq$0.57                    & $\geq$1.37                                           \\
120    & 16.944733      & -17.507444     & 6.2568                 & 6.70362              & $\geq$0.53                    & $\geq$0.94                                           \\
121    & 16.945259      & -17.508404     & 5.34374                & 6.97089              & $\geq$0.77                    & $\geq$0.95                                           \\
123    & 16.944564      & -17.506918     & 5.81343                & 6.01317              & $\geq$0.65                    & $\geq$0.86                                           \\
128    & 16.944503      & -17.506893     & 4.46098                & 6.70362              & $\geq$0.74                    & $\geq$0.82                                           \\
131    & 16.946019      & -17.510753     & 6.69745                & 8.06224              & 0.80                    & 9.72                                           \\
134    & 16.944898      & -17.50895      & 4.67393                & 6.70362              & 0.41                    & 0.41                                           \\
135    & 16.946266      & -17.511211     & 4.1396                 & 6.79271              & 0.50                    & 5.89                                           \\
140    & 16.94382       & -17.507018     & 5.29772                & 6.70362              & $\geq$0.48                    & $\geq$0.29                                           \\
141    & 16.943238      & -17.506891     & 4.46725                & 6.97089              & $\geq$0.53                    & $\geq$0.33                                           \\
142    & 16.943671      & -17.506675     & 5.37092                & 8.12908              & 0.69                    & 0.68                                           \\
144    & 16.944169      & -17.507668     & 6.12845                & 8.68588              & 0.81                    & 1.37                                           \\
145    & 16.944753      & -17.50903      & 5.27769                & 7.2159               & $\geq$1.05                    & $\geq$2.37                                           \\
150    & 16.943718      & -17.508348     & 4.74218                & 7.88407              & $\geq$0.56                    & $\geq$0.32                                           \\
153    & 16.94291       & -17.50806      & 5.36093                & 7.52771              & 0.47                    & 0.56                                           \\
154    & 16.943587      & -17.507753     & 4.7744                 & 6.70362              & 0.55                    & 0.53                                           \\
155    & 16.943661      & -17.508698     & 4.45091                & 6.70362              & $\geq$0.90                    & $\geq$0.88                                           \\
156    & 16.943324      & -17.508057     & 6.56955                & 6.70362              & $\geq$0.42                    & $\geq$0.29                                           \\
157    & 16.94353       & -17.508268     & 4.76702                & 7.23817              & 0.45                    & 0.28                                           \\
158    & 16.943514      & -17.508197     & 5.61672                & 8.03997              & 0.40                    & 0.25                                           \\
160    & 16.943091      & -17.507907     & 4.8025                 & 6.4809               & 0.59                    & 0.87                                           \\
162    & 16.943251      & -17.508129     & 5.81006                & 8.73044              & 1.50                    & 3.92                                           \\
179    & 16.9459361     & -17.5058213    & 5                      & 6                    & 0.57                    & 45.16 
\\
\hline
\multicolumn{4}{l}{\small {\bf{Note.}} $^a$ values taken from \cite{linden21}.} \\
\end{tabular}
\end{table*}

\end{document}